\documentclass[prd,10pt,aps,twocolumn,superscriptaddress,floatfix,notitlepage,nofootinbib,amssymb,amsmath]{revtex4-1}
\usepackage{mathtools}
\usepackage{epsfig}

\newcommand{\beqn}{\begin{eqnarray}}
\newcommand{\eeqn}{\end{eqnarray}}
\newcommand{\eq}[1]{(\ref{#1})}

\newcommand{\cS}{{\cal S}}

\newcommand{\cZ}{{\cal Z}}

\newcommand{\cD}{{\cal D}}

\newcommand{\Z}{{\mathbb Z}}
\newcommand{\bs}{\boldsymbol}
\newcommand{\plane}{{\cal P}_{\cal S}}
\newcommand{\avr}[1]{{\left\langle #1 \right\rangle}}

\begin{document}

\title{
The Casimir effect and deconfinement phase transition 
}

\author{M. N. Chernodub}
\affiliation{Laboratoire de Math\'ematiques et Physique Th\'eorique UMR 7350, Universit\'e de Tours, 37200 France}
\affiliation{Laboratory of Physics of Living Matter, Far Eastern Federal University, Sukhanova 8, Vladivostok, 690950, Russia}
\author{V. A. Goy}
\affiliation{Laboratory of Physics of Living Matter, Far Eastern Federal University, Sukhanova 8, Vladivostok, 690950, Russia}
\author{A. V. Molochkov}
\affiliation{Laboratory of Physics of Living Matter, Far Eastern Federal University, Sukhanova 8, Vladivostok, 690950, Russia}

\begin{abstract}
We show that the Casimir effect may lead to a deconfinement phase transition induced by the presence of boundaries in confining gauge theories. Using first-principle numerical simulations we demonstrate this phenomenon in the simplest case of the compact lattice electrodynamics in two spatial dimensions. We find that the critical temperature of the deconfinement transition in the vacuum between two parallel dielectric/metallic wires is a monotonically increasing function of the separation between the wires. At infinite separation the wires do not affect the critical temperature while at small separations the vacuum between the wires looses the confinement property due to modification of vacuum fluctuations of virtual monopoles.
\end{abstract}

\date{September 7, 2017}

\maketitle

\section{Introduction}

Quantum fluctuations of virtual particles are affected by the presence of physical objects. This property is a cornerstone of the Casimir effect~\cite{ref:Casimir} which predicts that the energy of vacuum (``zero-point'') quantum fluctuations is modified by the presence of physical bodies~\cite{ref:Bogdag,ref:Milton}. The energy shift of the virtual particles has real physical consequences because the Casimir effect leads to appearance a tiny force between neutral objects (called sometimes the ``Casimir-Polder force''~\cite{Casmir:1947hx}) which is detectable experimentally~\cite{ref:experiment}. The Casimir effect is an important phenomenon because it demonstrates the physical significance of the vacuum energy. 

The Casimir effect has been mostly studied in noninteracting field theories. However, apart from the simplest geometry of two parallel perfectly conducting plates, the calculation of the shift of the vacuum energy is a nontrivial analytical problem even for physical bodies of simplest geometries. Moreover, even in a free field theory the Casimir problem cannot be solved exactly for a physical object of an arbitrary shape. Therefore the Casimir effect is often studied using certain analytical approximations such as the proximity-force approximation~\cite{ref:proximity} as well as utilizing numerical tools~\cite{Johnson:2010ug} which includes worldline approaches~\cite{Gies:2006cq} and methods of lattice field theories~\cite{ref:Oleg,ref:paper:1,ref:paper:2}.

The Casimir energy is known to be modified in the presence of (self) interactions of the fields. In phenomenologically interesting case of quantum electrodynamics the correction to the Casimir-Polder forces comes from fermionic vacuum loops. Due to the weakness of the electromagnetic interaction, the corresponding correction may be calculated in the standard perturbation theory. The result, given by the second order perturbation theory, turns out to be negligibly small~\cite{Bordag:1983zk,ref:Milton}.

In strongly coupled theories the interactions may not only lead so a substantial modification of the Casimir force, but they may also affect the nonperturbative structure of the vacuum itself. For example, the Casimir effect in the mentioned double-plate geometry leads to strengthening of the finite-temperature phase transition associated with chiral symmetry breaking in a four-fermion effective field theory~\cite{Flachi:2013bc}. The boundary effects restore the chiral symmetry in a chirally broken phase both in plane~\cite{Tiburzi:2013vza} and in cylindrical~\cite{Chernodub:2016kxh} geometries.  The critical temperature of the restoration of the chiral symmetry depends on the geometry of the system. The interactions may also change the sign of the Casimir--Polder force in fermionic systems with condensates~\cite{Flachi:2017cdo} as well as in the $\mathbb{C}P^{N-1}$ model on an interval with the Dirichlet boundary conditions~\cite{Flachi:2017xat} (see also Ref.~\cite{Betti:2017zcm} for a related discussion).

In our paper we study the influence of the Casimir effect on a (de)confining phase transition in compact electrodynamics which is one of the simplest field theories possessing the linear confinement property. We use a general numerical approach which is applicable to a wide class of gauge theories including non-Abelian (Yang-Mills) gauge theory as well as various theories with interacting matter fields~\cite{ref:paper:1,ref:paper:2}. This numerical method, which is based on well-developed algorithms of lattice quantum (field) theory, allows us to investigate the Casimir effect from the first principles of the theory in a non-perturbative regime which is otherwise not accessible by standard perturbative calculations. 

In Refs.~\cite{ref:paper:1,ref:paper:2} we have studied Casimir energy induced by dielectric and perfectly metallic plates in the vacuum of the compact electrodynamics in two spatial dimensions. Since this model has both perturbative (photon) and nonperturbative (monopole) sectors, the plates not only modify the fluctuations of perturbative photon fields~\cite{ref:paper:1}, but they also affect the dynamics of the topological defects in the photon fields, i.e. the magnetic monopoles~\cite{ref:paper:2}. In particular, the ideal metallic plates cannot be pierced by the magnetic field coming from the monopoles which leads to the inhibition of the monopole gas between closely spaced metallic plates. The latter effect modifies strongly the Casimir energy of vacuum fluctuations and, consequently, affects nonperturbatively the Casimir-Polder force between the plates~\cite{ref:paper:2}. In the present paper we demonstrate the existence of a phenomenon which works in exactly the opposite way: the Casimir effect influences the vacuum structure of the theory by modifying the monopole dynamics which, in turn, induces deconfinement of electric charges in between closely spaced plates. 

The structure of this paper is as follows. In Sec.~\ref{sec:formulation} we briefly describe the lattice formulation of the Casimir effect following Refs.~\cite{ref:paper:1,ref:paper:2} and focusing on the compact electrodynamics. We present our numerical results on the monopole density in the finite temperature theory in Sec.~\ref{sec:monopole}. We discuss the effect of the presence of the dielectric wires on the monopoles and compare our numerical findings with the zero-temperature results of Ref.~\cite{ref:paper:2}. Section~\ref{sec:confinement} describes the outcomes of our numerical simulations on the confinement-deconfinement phase transition in the presence of dielectric/ideal-metallic wires. We also present the corresponding phase diagram of the model in the ``temperature -- interwire separation'' plane of parameters. Our conclusions on the role of the finite Casimir geometry in the deconfinement phase transition are summarized in the last section.

\section{Casimir effect on the lattice}
\label{sec:formulation}

In this section we briefly summarize the formulation of the Casimir problem for the compact lattice electrodynamics. More detailed discussion, including generalizations a wide class of other lattice (gauge) theories, can be found in Refs.~\cite{ref:paper:1,ref:paper:2}.

\subsection{Compact electrodynamics: a brief introduction}

In our paper we study the Casimir effect in the vacuum of compact electrodynamics\footnote{We would like to stress that the discussed model is often called ``the compact electrodynamics'' or ``the compact QED'', despite it contains no dynamical matter fields. In our paper we will use this standard terminology.} which has interesting nonperturbative features such as the linear confinement of electric charges and the mass gap generation in physically relevant cases of two and three spatial dimensions. 
The case of two spatial dimensions, considered in this paper, is particularly attractive since it can be treated using simple analytical techniques~\cite{Polyakov:1976fu}.
The linear confinement means that the long-ranged potential between the oppositely charged electric particles is linearly proportional to the distance between the particles $L$,
\beqn
V(L) = \sigma L\,,
\label{eq:V:L}
\eeqn
where $\sigma$ plays a role of the tension of the string which spans between the static charges and confines them together. Indeed, due to the linear nature of the inter-particle potential~\eq{eq:V:L} the particles can never be separated from each other at an infinite distance with expense of a finite amount of energy. 

In the confining phase, the model exhibits the phenomenon of the mass gap generation which implies that the mass of a lightest particle (photon, in our case) is nonzero~\cite{Polyakov:1976fu}.

In addition to the linear confinement and the mass gap properties, the model possesses a nontrivial phase diagram featuring a (deconfinement) phase transition at finite temperature. In the deconfinement phase the string tension $\sigma$ in Eq.~\eq{eq:V:L} vanishes and the electric charges are liberated. These features make the compact electrodynamics similar to (and, simultaneously, a simplified/toy model of) the Yang-Mills theory which is the most important nonperturbative ingredient of Quantum Chromodynamics.

The nonperturbative physics of the compact electrodynamics appears as a result of particular dynamics of Abelian monopoles which, in turn, emerge in this model as a result of the compactness of its Abelian gauge group. The monopoles are topological defects which are particle-like (instanton-like) objects in three (two) spatial dimensions. Basically, the model describes the dynamics of two types of physical entities: photons and monopoles. The photons govern the usual perturbative physics  while the monopoles are responsible for various nonperturbative effects. The former describe, for example, the Coulomb part of interaction potential between test electric charges while the latter lead to emergence of the linear confining potential between the charged particles, the mass gap generation and the associated phase transitions. 

The compact electrodynamics has the Abelian gauge group $U(1)$ which is very similar to the one of the usual non-compact electrodynamics. Thus both theories possess the photons in their physical spectra. However, the compact electrodynamics has also the Abelian monopoles which are absent in the standard noncompact version of the model. This model may be viewed as a long-range (infrared) effective model of more complicated non-Abelian gauge theories which possess monopoles as natural topological excitations (one may mention the 't~Hooft--Polyakov monopoles~\cite{tHooft:1974kcl,Polyakov:1974ek} in the Georgi-Glashow model~\cite{Georgi:1974sy} as a simplest example). These monopoles are essentially non-Abelian objects in the vicinity of the monopole center (in the monopole core) while outside the core the monopoles are Abelian objects with Abelian magnetic field characterized by nonzero (quantized) magnetic charge. Since the pure Abelian monopole is characterized by a singularity (``a defect'') of the photon field in its geometrical center, it is often said that the monopoles are the topological defects in the theory. In the case of non-Abelian monopoles such as the 't~Hooft--Polyakov monopole~\cite{tHooft:1974kcl,Polyakov:1974ek}, the singularity in the monopole core is softened by other non-Abelian (matter) fields.

The compact QED is a toy model not only for particle physics, but it is also serves as an effective macroscopic model in a number of condensed matter systems~\cite{ref:book:Herbut}. Despite its apparent complexity, the compact QED may be easily formulated both in continuum spacetime~\cite{ref:book:Kleinert} convenient for analytical calculations as well as in discretized (lattice) spacetimes suitable for numerical simulations. The model in two space dimensions is advantageous from the point of view of both numerical and analytical calculations, in particular, because of the technique developed in Ref.~\cite{Polyakov:1976fu}. In our paper we restrict ourselves to (2+1) dimensional spacetime dimensions where the Casimir plates are, in fact, wires due to reduced dimensionality (sometimes we continue to call them as ``the plates'', though). We study the model in the lattice regularization which is convenient for numerical simulations. 

\subsection{Compact electrodynamics on the lattice}

The compact lattice electrodynamics in three Euclidean dimensions is determined by the following action
\beqn
S[\theta] = \beta \sum_P \left(1 - \cos \theta_P \right)\,,
\label{eq:S}
\eeqn
where the sum is taken over all elementary plaquettes $P \equiv P_{x,\mu\nu}$ of the lattice. Each plaquette is characterized by the position~$x$ of one of its corners and by orientation in the plaquette plane determined by two orthogonal vectors $\mu < \nu$ with $\mu,\nu = 1, 2, 3$. The lattice gauge field $\theta_{x,\mu} \in [-\pi,+\pi)$ is a compact dynamical variable defined at each link $l_{x,\mu}$ starting at the point $x$ and pointing in the direction $\mu$.  The dimensionless lattice angle $\theta_{x,\mu}$ is related to the dimensionful continuum gauge field $A_\mu(x)$ as follows: $\theta_{x\mu} = a A_{\mu}(x)$, where $a$ is the lattice spacing (the physical length of elementary link of the lattice). The lattice action~\eq{eq:S} is expressed via the plaquette angles
\beqn
\theta_{P_{x,\mu\nu}} = \theta_{x,\mu} + \theta_{x+\hat\mu,\nu} - \theta_{x+\hat\nu,\mu} - \theta_{x,\nu}\,,
\label{eq:theta:P}
\eeqn
which play the role of the lattice field strength. In the continuum limit, $a\to 0$, the plaquette variable~\eq{eq:theta:P} tends to its continuum version $\theta_{P_{x,\mu\nu}} = a^2 F_{\mu\nu}(x) + O(a^4)$ for small fluctuations of the photon fields. Here $F_{\mu\nu} = \partial_\mu A_\nu - \partial_\nu A_\mu$ is the field strength tensor of the electromagnetic field in continuous (non-discretized) spacetime. 

Consequently, the lattice action~\eq{eq:S} becomes the standard photon action for weak gauge fields $A_\mu$, if one associates the lattice coupling constant with the lattice spacing $a$ and the electric charge $g$:
\beqn
\beta = \frac{1}{g^2 a}\,.
\label{eq:beta:a}
\eeqn
In three space-time dimensions the continuum gauge coupling $g$ is a dimensionful quantity $[g] = {\text{mass}}^{1/2}$. The relation~\eq{eq:beta:a} is valid in the regime of the weak coupling~$g$ corresponding to large values of the lattice couplings~$\beta$. In this regime the expansion of the lattice action in terms of small fluctuations of the gauge field provides us with a link between the lattice and continuum versions of this model.

In the presence of monopole singularities the continuum action becomes more complicated as it includes singular Dirac lines attached to the Abelian monopoles. We discuss the continuum formulation of compact QED in our previous paper~\cite{ref:paper:2}. A more detailed review on compact (gauge) fields and topological object can be found in the book~\cite{ref:book:Kleinert}.

\subsection{Compactness and monopoles}

The action of the model~\eq{eq:S} is invariant under discrete shifts of the plaquette variable, $\theta_P \to \theta_P + 2 \pi n$ where $n$ is an arbitrary integer number, $n \in \Z$. This invariance implies that two lattice field strengths $\theta_P$ and $\theta_P' = \theta_P + 2 \pi n$ with $n \in \Z$ are physically equivalent to each other. Therefore the Abelian gauge group of the theory corresponds to a compact manifold (hence the name ``compact'' electrodynamics).

In the continuum limit the $2 \pi$ shifts of the action lead to singularities in the continuum version of the field strength~\eq{eq:theta:P} as they generate contributions of the form $\delta F_{\mu\mu} \sim 2 \pi/a^2$ where $a \to 0$ is the vanishing lattice spacing. In four space-time dimensions these shifts correspond to gauge-dependent displacements of the Dirac sheets which are worldlines of thin Dirac strings attached to the Abelian monopoles. The ends of the Dirac strings correspond to positions of the Abelian monopoles which are physical, gauge-invariant topological defects. Thus, the compactness of the model leads to the appearance of singular configurations of gauge fields, the Abelian monopoles.

In the (2+1)-dimensional compact electrodynamics the monopoles are instanton-like objects. In the lattice formulation of the theory the monopole density is defined as the  divergence of the physical part ${\bar \theta}_P$ of the lattice field-strength tensor $\theta_P$ at three-dimensional cubes $C_x$:
\beqn
\rho_x = \frac{1}{2\pi} \sum_{P \partial C_x} {\bar \theta}_P\,,
\label{eq:rho:lattice}
\eeqn
where
\beqn
{\bar \theta}_P = \theta_P + 2 \pi k_P \in [-\pi,\pi), \qquad k_P \in \Z.
\label{eq:bar:theta}
\eeqn
Here the integer number $k_P$ is chosen in such a way that the physical plaquette angle ${\bar \theta}_P$ is limited within the interval $[-\pi,\pi)$. In the continuum limit Eq.~\eq{eq:rho:lattice} is proportional to the divergence of the magnetic field. 

The lattice monopole density~\eq{eq:rho:lattice} at each cube $C_x$ is always an integer number, $\rho_x \in \Z$, so that the quantity~\eq{eq:rho:lattice} maybe associated with a number of continuum Abelian monopoles inside each lattice cube $C_x$. The monopoles were studied intensively both in Abelian and non-Abelian lattice gauge theories~\cite{Chernodub:1997ay}.

\subsection{Casimir boundary conditions on the lattice}

Similarly to case of the gauge theories in continuum spacetime, the Casimir boundary conditions may also be formulated in the discretized space, both for Abelian and non-Abelian lattice gauge theories~\cite{ref:paper:1} . Here we briefly describe, following Ref.~\cite{ref:paper:2}, the main statements which are relevant to the compact electrodynamics studied in this paper.

In (3+1) dimensions the Casimir problem is defined for two-dimensional surfaces of physical materials. If the surfaces are made of an ideal metal, then two tangential (with respect to the surface at each local point) components of the electric field and a normal component of the magnetic field vanish. These requirements lead to modification of zero-point (vacuum) fluctuations and appearance of forces between the uncharged surfaces.

In (2+1) dimensions the Casimir boundary conditions are applied to one dimensional wires. In the case of ideal metal a tangential component of the electric field is forced to vanish at each wire. In a covariant form the corresponding boundary conditions read as follows:
\beqn
F^{\mu\nu}(x) s_{\mu\nu}(x) = 0\,,
\label{eq:F:0:cov}
\eeqn
where $F_{\mu\nu} = \partial_{[\mu,} A_{\nu]} \equiv \partial_\mu A_\nu - \partial_\nu A_\mu$ is the field strength tensor and  
\beqn
s_{\mu\nu}(x) = \int d \tau \int d \xi \, \frac{\partial {\bar x}_{[\mu,}}{\partial \tau} \frac{\partial {\bar x}_{\nu]}}{\partial \xi} \,
\delta^{(3)}\bigl(x - {\bar x}(\tau,\xi)\bigr), \quad
\label{eq:s:munu:gen}
\eeqn
is the local surface element of the world sheet of the wire. The latter is described by the vector function ${\bar x}_\mu = {\bar x}_\mu(\tau,\xi)$ parameterized by two parameters $\tau$ and~$\xi$.

In this paper we consider two static straight wires directed along the $x_2$ axis and separated along the $x_1$ direction at $x_1 = l_1$ and $x_1 = l_2$. For each such wire, the local surface element of the corresponding world sheet~\eq{eq:s:munu:gen} is $s_{\mu\nu}(x) = (\delta_{\mu, 2} \delta_{\mu, 3} - \delta_{\mu, 3} \delta_{\mu, 2}) \delta(x_3 - l_a)$ where the parameter $a=1,2$ labels the wires and the $x_3$ axis is associated with the Euclidean ``time'' direction. Consequently, the covariant condition~\eq{eq:F:0:cov},
\beqn
F^{23}(x) {\biggl|}_{x_1 = l_a} = 0, \qquad a = 1,2\,,
\label{eq:F23}
\eeqn
naturally forces the tangential component of the electric field to vanish at each wire. In the lattice gauge theory this boundary condition corresponds to the vanishing of the field strength tensor~\eq{eq:theta:P} -- up to the discrete compact transformations mentioned above -- at a set of the plaquettes $P \in \plane$ that belongs to the world surfaces~$\cS$ of the wires.

In the case of an ideal metal, the Casimir boundary condition is given by the lattice version of Eq.~\eq{eq:F23}:
\beqn
\cos\theta_{x,23} {\biggl|}_{x_1 = l_a} = 1, \qquad a = 1,2\,,
\label{eq:F01:latt:3D}
\eeqn
for all possible $x_2$ and $x_3$ and fixed $x_1 = l_1$ or $x_1 = l_2$.

A simplest way to implement the boundary condition~\eq{eq:F01:latt:3D} is to add a set of Lagrange multipliers to the standard Abelian action~\cite{ref:paper:1,ref:paper:2}: 
\beqn
S_{\varepsilon}[\theta;\plane] = \sum_P \beta_P(\varepsilon) \cos \theta_P\,,
\label{eq:S:beta}
\eeqn
where the plaquette-dependent gauge coupling:
\beqn
\beta_{P_{x,\mu\nu}} (\varepsilon) = \beta \bigl[1 + (\varepsilon - 1)\, & & (\delta_{\mu,2} \delta_{\nu,3} + \delta_{\mu,3} \delta_{\nu,2}) \nonumber \\
& & \cdot \left(\delta_{x,l_1} + \delta_{x,l_2}\right)\bigr]\,.
\label{eq:beta:P:3d}
\eeqn
is a function of the dielectric permittivity $\varepsilon$ of the wire. At $\varepsilon = 1$ the wires are absent and we get the original theory. In the limit $\varepsilon \to \infty$ the components of the physical lattice field-strength tensor~\eq{eq:bar:theta} vanish at the word-surfaces of the wires, ${\bar \theta}_{x,23} = 0$, as required by the lattice condition~\eq{eq:F01:latt:3D}.

In addition to the dielectric permittivity $\varepsilon$, it is also convenient to characterize the properties of the Casimir wires in two space dimensions in terms of the  ``relative strength'' of the wires:
\beqn
\delta \beta = \beta_{\mathrm{wire}} - \beta \equiv (\varepsilon - 1) \beta
\label{eq:delta:beta}
\eeqn
which is related to the permittivity $\varepsilon$ of the wires using the standard formula
\beqn
\varepsilon = 1 + \chi_\varepsilon\,,
\label{eq:permittivity:chi}
\eeqn
where
\beqn
\chi_\varepsilon = \frac{\delta \beta}{\beta}\,,
\label{eq:chi}
\eeqn
is the conventional electric susceptibility of the wire material.

Physically, the wire strength $\delta \beta$ characterizes the degree of influence of the wire on the fluctuations of the electromagnetic field: at $\delta \beta = 0$ (equivalently, at $\varepsilon = 1$) the wire is absent while at $\delta \beta \to \infty$ (i.e., at $\varepsilon \to \infty$) the effect of the wire is the strongest possible and the wire become perfectly metallic~\cite{ref:Bogdag}. For the sake of simplicity, in our paper we study the Casimir effect in between two wires of the equal strength~\eq{eq:delta:beta}. 

In the absence of the Casimir boundaries the partition function of compact electrodynamics includes the integration over the angular link variables $\theta_{x,\mu}$:
\beqn
\cZ = \prod_{l} \int_{-\pi}^\pi d \theta_l \, e^{-S[\theta]}\,,
\label{eq:Z:lat}
\eeqn
where the Abelian action $S$ is given in Eq.~\eq{eq:S}. In the presence of the Casimir plates made of the ideal metal the partition function~\eq{eq:Z:lat} is modified
\beqn
\cZ[\plane] & = & \lim_{\varepsilon \to + \infty} \cZ_\varepsilon[\plane]\,, 
\label{eq:Z:Casimir:1}
\eeqn
where the partition function
\beqn
\cZ_\varepsilon[\plane] & = & \int \cD \theta \, e^{-S_\varepsilon[\theta;\plane]}\,.
\label{eq:Z:Casimir:2}
\eeqn
corresponds to the wires of finite permittivity $\varepsilon$.

In our simulations we realize the case of perfectly conducting wires by taking the limit~\eq{eq:Z:Casimir:1} of large permittivity $\varepsilon \to \infty$ in which a component of the electric field parallel to the wire vanishes~\eq{eq:F23} thus mimicking an ideal metal. We would like also to stress that in two spatial dimensions the magnetic permeability does not exist and a wire with infinite static dielectric permittivity affects the electromagnetic field in the same way as an ideal metal (we refer a reader to Section 5.1 of Ref.~\cite{ref:Bogdag} where this limit is discussed in more details).

\subsection{Details of numerical calculations}

We consider the model at asymmetric cubic lattice $L_t \times L_s^2$ with periodic boundary conditions along all three directions of the Euclidean lattice. In our simulations we take $L_t = 4$ and $L_s = 64 \gg L_t$. We also set two parallel static straight wires with flat parallel time-sheets in the $(x_2,x_3)$ planes at the positions $x_1 = l_1$ and $x_1 = l_2$. The lattice extension in the Euclidean time (or, inverse temperature) direction $L_t$ is related to the physical temperature of thermal equilibrium
\beqn
T = \frac{1}{L_t a}\,,
\label{eq:T:a}
\eeqn
where $a$ is the lattice spacing related to the physical gauge coupling $g$ and lattice gauge coupling $\beta$ by Eq.~\eq{eq:beta:a}.

The Casimir wires are implemented with the help of the space-dependent gauge coupling~\eq{eq:beta:P:3d} in the action of the model~\eq{eq:S:beta}. Due to periodic boundary conditions the wires divide the $x_1$ axis into two, generally inequivalent, intervals, $R$ and $L_s - R$, where $R = |l_2 - l_1| \leqslant L_s/2$ is the shortest distance between the wires.  Due to the boundary conditions all $R$-dependent quantities are invariant under the spatial flip in the $x_1$ direction, $R \to L_s - R$. The wire strengths~\eq{eq:delta:beta} are taken in the wide range $\delta \beta = 0.5 \dots 88$. 

Our basic simulation parameters are the same as the ones in our previous simulations~\cite{ref:paper:1,ref:paper:2}. All configurations of the gauge field are generated with the help of a Hybrid Monte Carlo algorithm based on standard Monte-Carlo methods improved by molecular dynamics algorithms~\cite{ref:Gattringer}. The latter utilize a second-order minimum norm integrator~\cite{ref:Omelyan} at several time scales~\cite{ref:Sexton}. The use of different timescales is a crucial tool which allows us to diminish integration errors accumulated at and outside worldsheets of the wires at which the Casimir boundary conditions are imposed. Long autocorrelation lengths in Markov chains are eliminated, following Ref.~\cite{ref:Gattringer}, with the help of five subsequent overrelaxation steps which separate gauge field configurations sufficiently far from each other.  We apply a self-tuning adaptive algorithm in order to control the acceptance rate of the Hybrid Monte-Carlo in a reasonable range between 0.70 and 0.85.  We use about $10^5$ trajectories per one value of the wire permittivity $\varepsilon$ with about $10^4$ trajectories for thermalization of our gauge configurations. 

\section{Casimir effect and monopoles}
\label{sec:monopole}

The presence of the Casimir (metallic or dielectric) plates affects both the fluctuations of the photon field and the dynamics of the monopoles between the worldsheets of wires (``plates''). The modification of the photon fields leads to the well-studied conventional Casimir effect while monopole dynamics in between the plates may also lead to a number of new nonperturbative modification of the Casimir effect which were first touched in Ref.~\cite{ref:paper:2}:

\begin{itemize}

\item[(i)] The presence of the monopoles influences the Casimir force in between the plates. Indeed, the presence of the monopoles leads to the generation of the mass gap which makes the photon massive, and, in turn, naturally suppresses the Casimir effect at large distances between the plates. This phenomenon was numerically found at zero temperature in Ref.~\cite{ref:paper:2}. 

\item[(ii)] The finite geometry of the Casimir setup modifies the dynamics of the monopoles in between the wires and should, consequently, affect the phase transition between the confined (mass-gapped) phase and the deconfined (gapless) phase of the model. This nonperturbative effect is the scope of the present paper and it will be discussed in details below. 

\end{itemize}

Since the discussion in this section has largely a qualitative nature, we work with the absolute monopole density defined in the natural lattice units, 
\beqn
\rho = \avr{|\rho_x|} \equiv \rho_{\mathrm{lat}}^{\mathrm{mon}} = \rho^{\mathrm{mon}}_{\mathrm{phys}} a^3,
\label{eq:rho:a3}
\eeqn
where the local monopole charge density~$\rho_x$ is given in Eq.~\eq{eq:rho:lattice}. In Eq.~\eq{eq:rho:a3} the lattice spacing $a$ is expressed via the gauge coupling $g$ and the lattice coupling $\beta$ using the simple formula~\eq{eq:beta:a}. 

\begin{figure}[!thb]
\begin{center}
\vskip 3mm
\includegraphics[scale=0.525,clip=true]{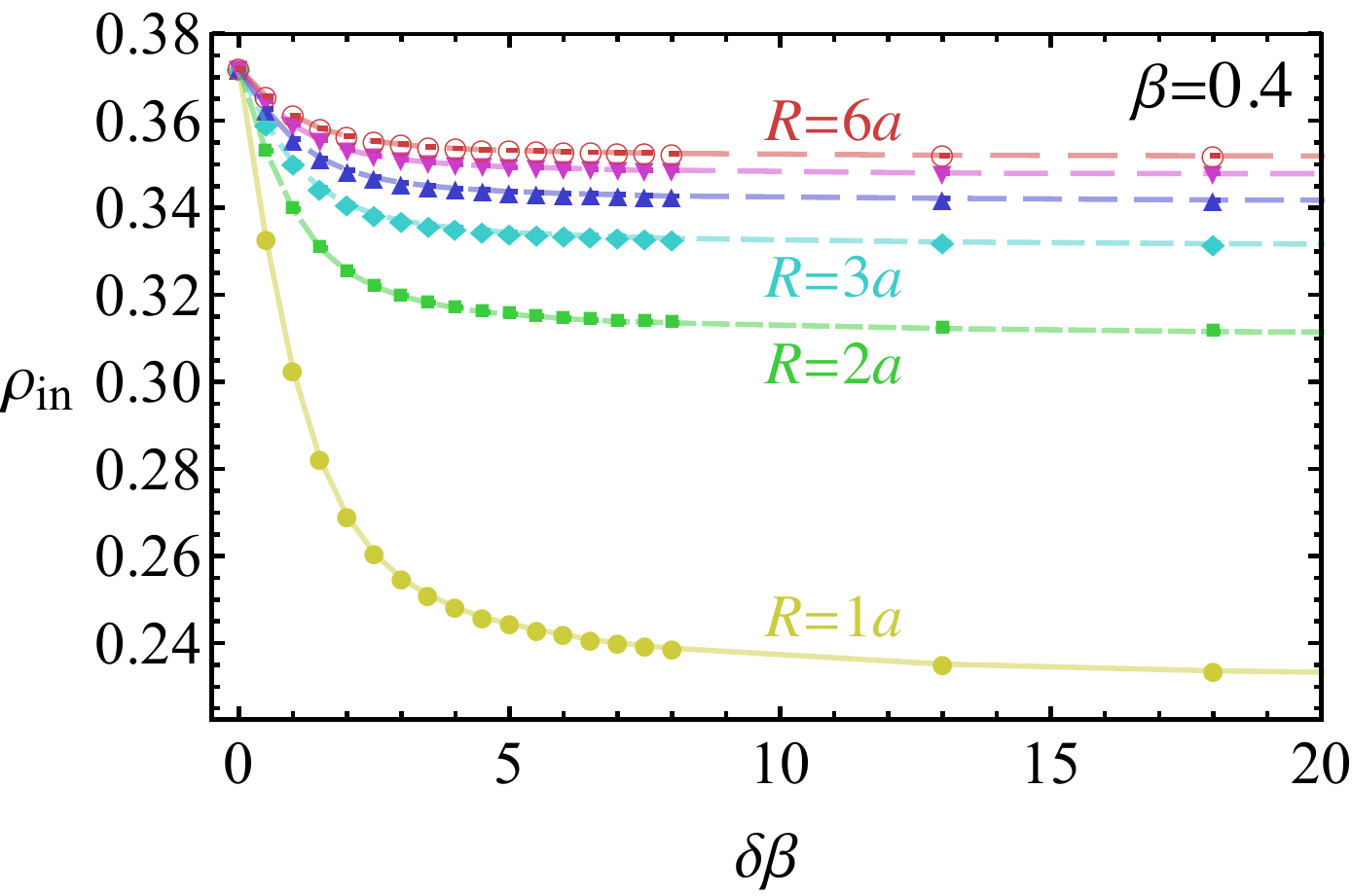}
\end{center}
\vskip -2mm 
\caption{The monopole density in lattice units~\eq{eq:rho:a3} as the function of the wire strength~\eq{eq:delta:beta} for various fixed separations between the plates $R$ at the lattice gauge coupling $\beta=0.4$. The error bars are smaller then the size of the markers. The lines are plotted to guide eyes.}
\label{fig:mon:vs:delta:beta}
\end{figure}

The simplest question we may ask ourselves is as follows: what is the effect of the Casimir wires on the monopole density in the space between them? In Fig.~\ref{fig:mon:vs:delta:beta} we plot the monopole density in between the wires as a function of the wire strength $\delta \beta$, defined in Eq.~\eq{eq:delta:beta}, for a certain set of fixed separations $R$ between the wires. We choose the small value of the lattice gauge coupling $\beta=0.4$ where the monopole density is high and the discussed effects are well pronounced.

Figure~\ref{fig:mon:vs:delta:beta} demonstrates the existence of two nonperturbative effects highlighting how the wires act on the monopoles. Firstly, the monopole density calculated in the space in between the wires~$\rho_{\mathrm{in}}$ turns ot to be a diminishing function of the wire strength $\delta \beta$ (or, equivalently, of the permittivity $\varepsilon$). This property is not unexpected since the metallic wires confine the monopole flux emanating by the monopoles to the narrow space between the wires. The squeezed monopole flux has bigger energy density compared to the energetically preferred spherical configuration of the magnetic flux around the monopole. Consequently, the plates make the associated monopole mass heavier. In turn, the heavier monopoles the lower their density. We come to the conclusion that the wires with bigger permittivity $\varepsilon$ (or, equivalently, the bigger wire strength $\delta\beta$) suppress the monopole density more effectively, in a full qualitative agreement with the behavior of the monopole density shown in Fig.~\ref{fig:mon:vs:delta:beta}.

Secondly, we see from Fig.~\ref{fig:mon:vs:delta:beta} that the wider the separation between the plates $R$ the higher the monopole density. This observation comes out naturally from the very same property: the smaller separation between the plates, the more squeezed the magnetic field, the higher its energy and, consequently, the heavier the monopole mass. More massive monopoles should have lower density compared to their lighter counterparts.  Thus, the smaller separation between the plates the more dilute the monopole gas and vice versa.

\begin{figure}[!thb]
\begin{center}
\vskip 3mm
\includegraphics[scale=0.525,clip=true]{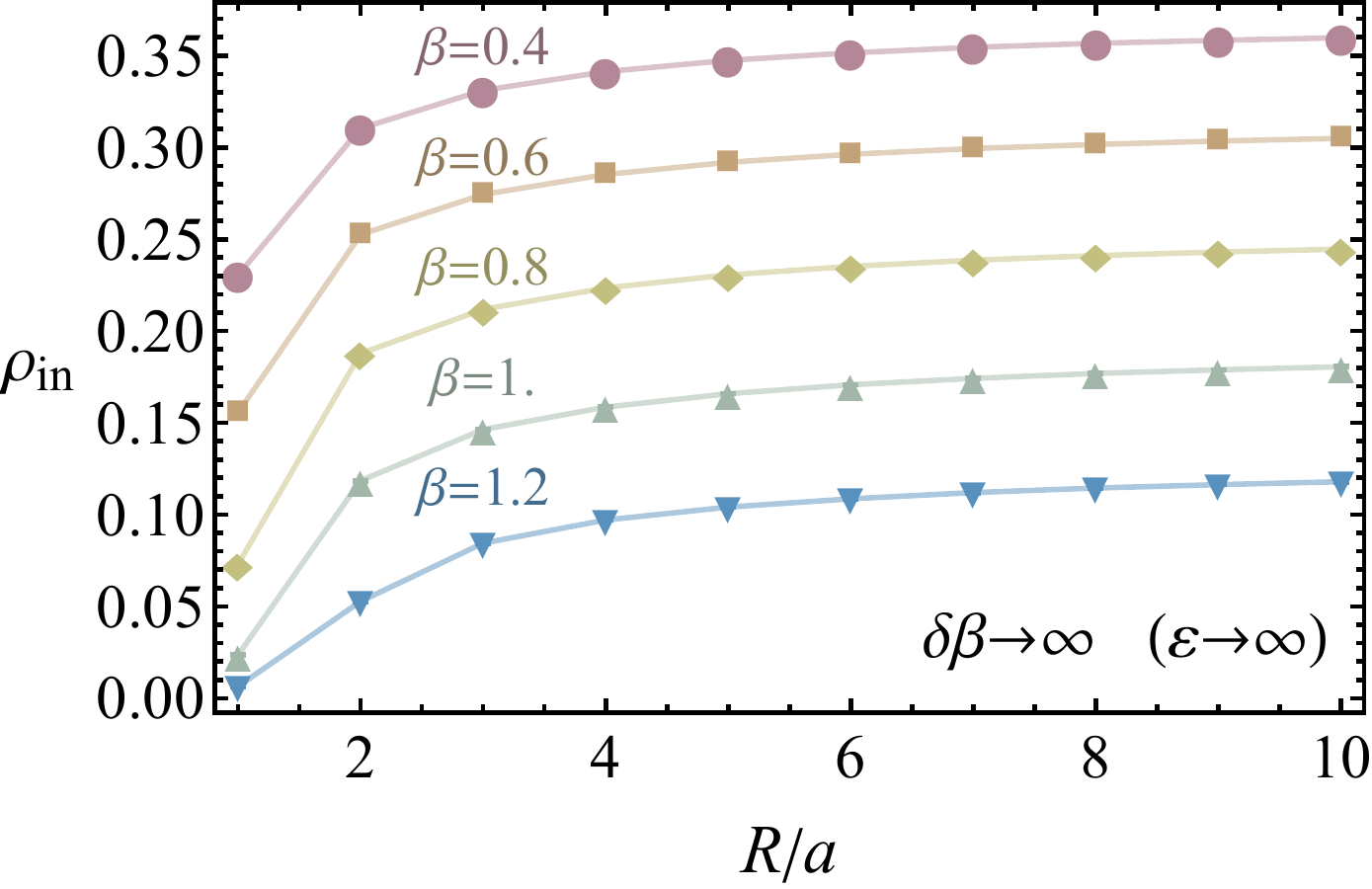}
\end{center}
\vskip -2mm 
\caption{Monopole density in between wires in the perfectly metallic limit $\varepsilon \to \infty$ as the function of interwire separation~$R$ for various fixed gauge couplings $\beta$. Notations are the same as in Fig.~\ref{fig:mon:vs:delta:beta}.}
\label{fig:mon:vs:R}
\end{figure}

In Fig.~\ref{fig:mon:vs:R} we plot the monopole density in between the plates in the perfect metallic limit ($\varepsilon \to \infty$) for various lattice couplings $\beta$. In agreement with the discussion above, the monopole density $\rho_{\mathrm{in}}$ is an increasing function of there interwire distance $R$: the further wires the lower they have the effect on the monopoles. As in the absence of the wires, the monopole density $\rho_{\mathrm{in}}$ is a decreasing function of the monopole gauge coupling $\beta$. Both Fig.~\ref{fig:mon:vs:delta:beta} and Fig.~\ref{fig:mon:vs:R} show self-consistently that the approaching (obviously, in terms of the interwire distance~$R$) and strengthening (in terms of the coupling difference~$\delta \beta$) wires suppress the monopole density between them.

The monopole density effects at finite temperature, shown both in Fig.~\ref{fig:mon:vs:delta:beta} and Fig.~\ref{fig:mon:vs:R}, agree qualitatively with the results of our previous studies performed at zero temperature~\cite{ref:paper:2}.

\begin{figure}[!thb]
\begin{center}
\vskip 3mm
\includegraphics[scale=0.525,clip=true]{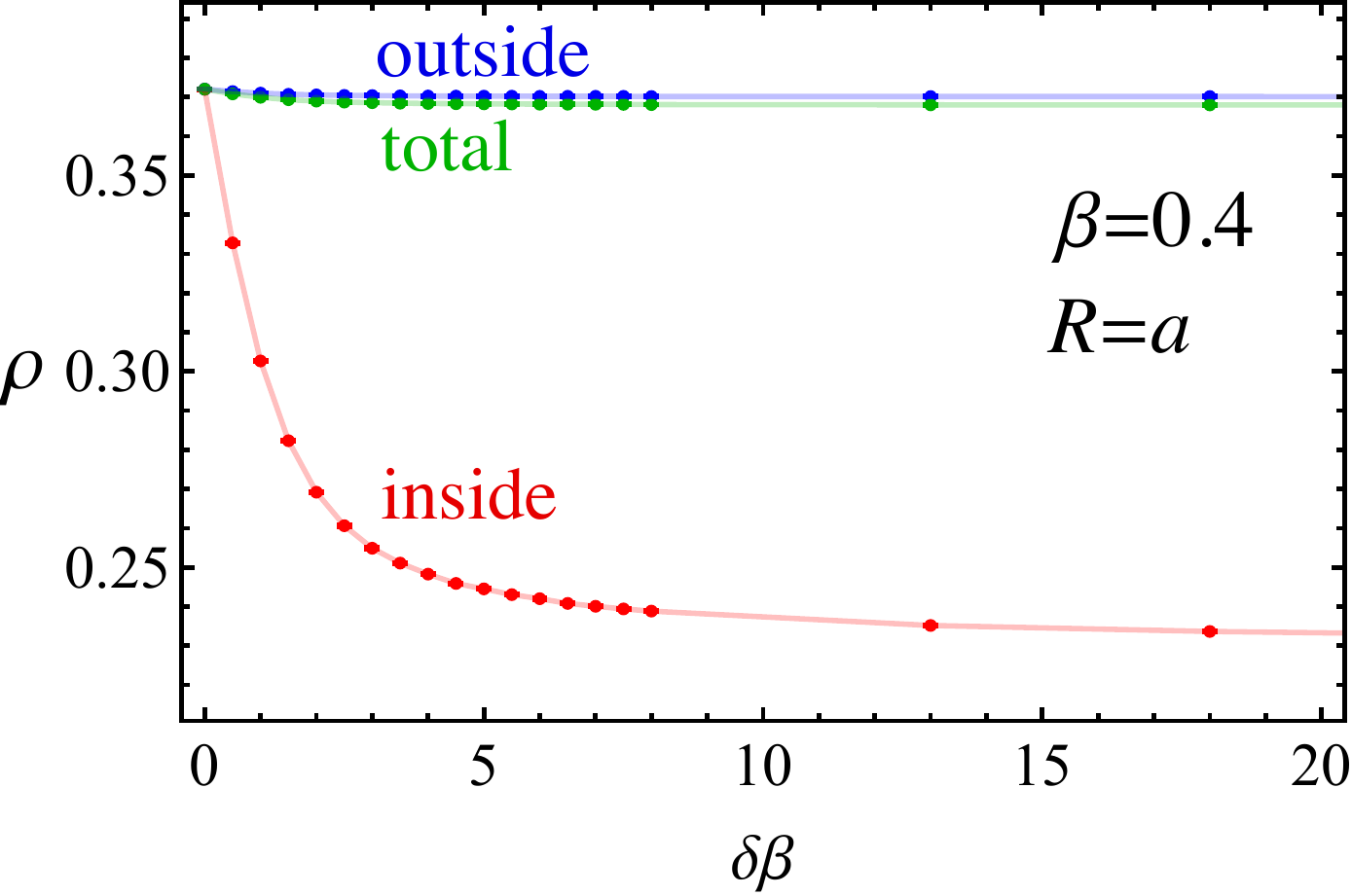}
\end{center}
\vskip -2mm 
\caption{Monopole density in between the plates (the red dots) and outside the plates (the blue dots) as compared to the total density of monopoles on the whole lattice (the green dots) at strong coupling $\beta=0.4$ and smallest separation between the Casimir wires, $R = a$.}
\label{fig:mon:compare}
\end{figure}

Finally, we would like to compare the effect of the Casimir wires on the relative density of the monopoles: how strong is the effect of the wires in the monopole density in between the wires (``inside'')  and in the outer space (``outside''). In order to quantify the effect of the wires, we show the mentioned quantities, as well as the total density of the monopoles, in Fig.~\ref{fig:mon:compare}. In order to highlight the magnitude of the effects we work at the smallest lattice gauge coupling $\beta=0.4$ (at which the monopole density is the largest) and the smallest separation between the plates ($R=a$), at which the effect of the latter is the strongest. Figure~\ref{fig:mon:compare} shows us that, in agreement with our intuitive expectations supported by Fig.~\ref{fig:mon:vs:delta:beta} and Fig.~\ref{fig:mon:vs:R}, the density of the monopoles in between the plates is a rapidly diminishing function of the plate strength $\delta \beta$ which has, however, a definite limit at $\delta \beta \to \infty$. At the same time the effect of the plates on the density of monopoles outside the plates, and on the total monopole density, is vanishingly small. Basically, the plates effectively expel the monopoles from the space between them to the space outside the plates as the density of the monopoles outside the plates is slightly larger compared to the total density of the monopoles. Thus, the plates affect only the properties of the vacuum in between (and not outside) them, as expected.

Summarizing, we have observed the suppression effect of the wires on the monopole density in the space between them. This observation suggests that the wires should suppress the confining properties of the vacuum. The effect of the Casimir geometry on confinement is discussed in the next section.

\section{Deconfinement phase transition}
\label{sec:confinement}

\subsection{Polyakov line as the order parameter}

The confining properties of the vacuum may be characterized by the Polyakov line
\beqn
L_{\bs x}(\theta) = \exp\left\{i \sum_{x_3 = 0}^{L_t -1} \theta_{x,3}\right\},
\label{eq:L:x}
\eeqn
where the sum of the time component ($\mu=3$) of the vector gauge field $\theta_{x,\mu} \equiv \theta_\mu(x)$ is taken along the Euclidean time direction $\tau \equiv x_3$. By the construction, the Polyakov loop $L_{\bs x}$ is a spatially local operator which is defined at a spatial point ${\bs x} = (x_1,x_2)$ and which does not depend on the (Euclidean) time coordinate $x_3$.

The vacuum expectation value of the Polyakov line~\eq{eq:L:x} is an order parameter of the deconfinement phase transition: the Polyakov line $\avr{L_{\bs x}}$ is vanishing in the confinement phase and it is nonzero in the deconfinement phase. The expectation value of the line operator $L_{\bs x}$ is associated with the free energy $F_{\bs x}$ of an isolated static electric charge localized at the point ${\bs x}$:
\beqn
e^{- F_{\bs x}/T} = \avr{L_{\bs x}}\,,
\label{eq:Fx:definition}
\eeqn
where $T$ is the temperature of the system. According to Eqs.~\eq{eq:beta:a} and \eq{eq:T:a} the physical temperature $T$, expressed in units of the coupling constant $g^2$, is a linear function of the lattice gauge coupling $\beta$:
\beqn
\frac{T}{g^2} = \frac{\beta}{L_t}\,,
\label{eq:T:phys}
\eeqn

In the confinement phase (low $T$ and small $\beta$) the order parameter $\avr{L_{\bs x}}$ is zero, implying that the free energy $F_{\bs x}$ is infinite, so that an isolated electric charge cannot exist. In the deconfinement phase (high $T$ and large $\beta$) both the order parameter and the associated energy are nonzero which is a clear indication of the existence of free electric charges (deconfinement).

In the presence of the plates it is convenient to identity three types of bulk expectation values of the Polyakov loop. They correspond to the expectation value taken over the whole space ($L_{\mathrm{tot}}$), the space in between (``inside'') the plates ($L_{\mathrm{in}}$) and the space outside the plates ($L_{\mathrm{out}}$), respectively:
\beqn
L_{\mathrm{tot}} & =&  \frac{1}{L_s^2} \avr{\sum_{x_1 = 0}^{L_s - 1} \sum_{x_2 = 0}^{L_s - 1} L_{x_1,x_2}}\,, 
\label{eq:L:tot}
\\
L_{\mathrm{in}} & =&  \frac{1}{L_s (R-1)} \avr{\sum_{x_1 = l_1 + 1}^{l_2 - 1} \sum_{x_2 = 0}^{L_s - 1} L_{x_1,x_2}}\,, 
\label{eq:L:in}
\\
L_{\mathrm{out}} & =&  \frac{1}{L_s (L_s - R-1)} \nonumber \\
& & \cdot \avr{\left(\sum_{x_1 = 0}^{l_1 - 1} + \sum_{x_1 = l_2+1}^{L_s - 1} \right)\sum_{x_2 = 0}^{L_s - 1}  L_{x_1,x_2}}\,.
\label{eq:L:out}
\eeqn
Here $L_s$ is the spatial extent of the lattice (in our calculations $L_s = 64$ in both directions $x_1$ and $x_2$) and $R = |l_1 - l_2| \leqslant L/2$ is the shortest distance between the plates (notice that as we consider the periodic boundary conditions so that the distance between the plates may be defined both as $R$ and $L_s - R$). The inner and outer expectation values of the Polyakov line, given in Eqs.~\eq{eq:L:in} and~\eq{eq:L:out} respectively, do not include contributions from the points belonging to the plates themselves (i.e., with $x_1 \neq l_{1,2}$). Therefore the average of the Polyakov line in between the plates~\eq{eq:L:in} is well defined only for separations $R \geqslant 2$ (for the sake of simplicity, we define the distances in dimensionless lattice units by setting $a=1$ in these formulae). 

\begin{figure}[!thb]
\begin{center}
\vskip 3mm
\includegraphics[scale=0.55,clip=true]{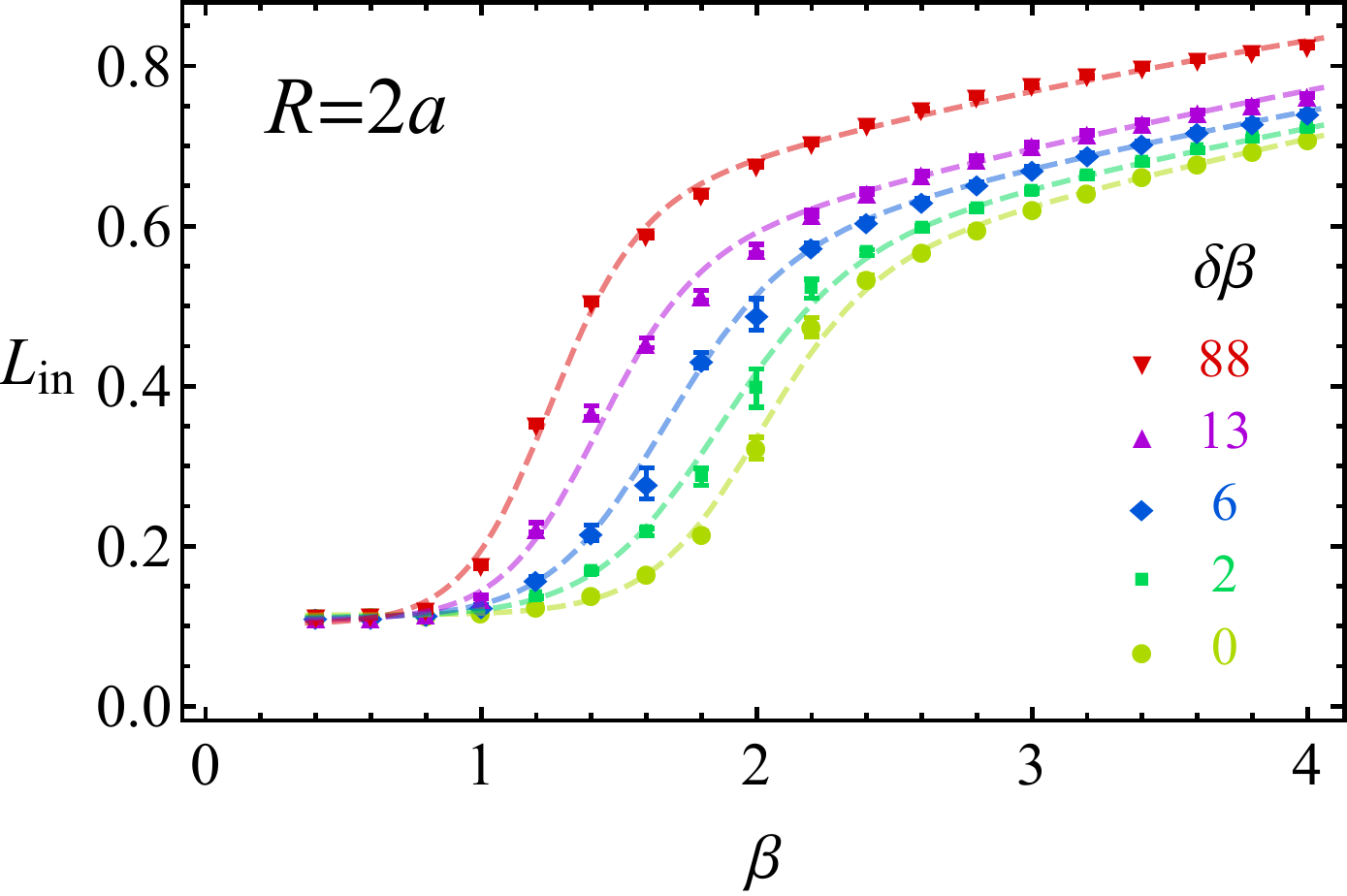} \\[5mm]
\includegraphics[scale=0.55,clip=true]{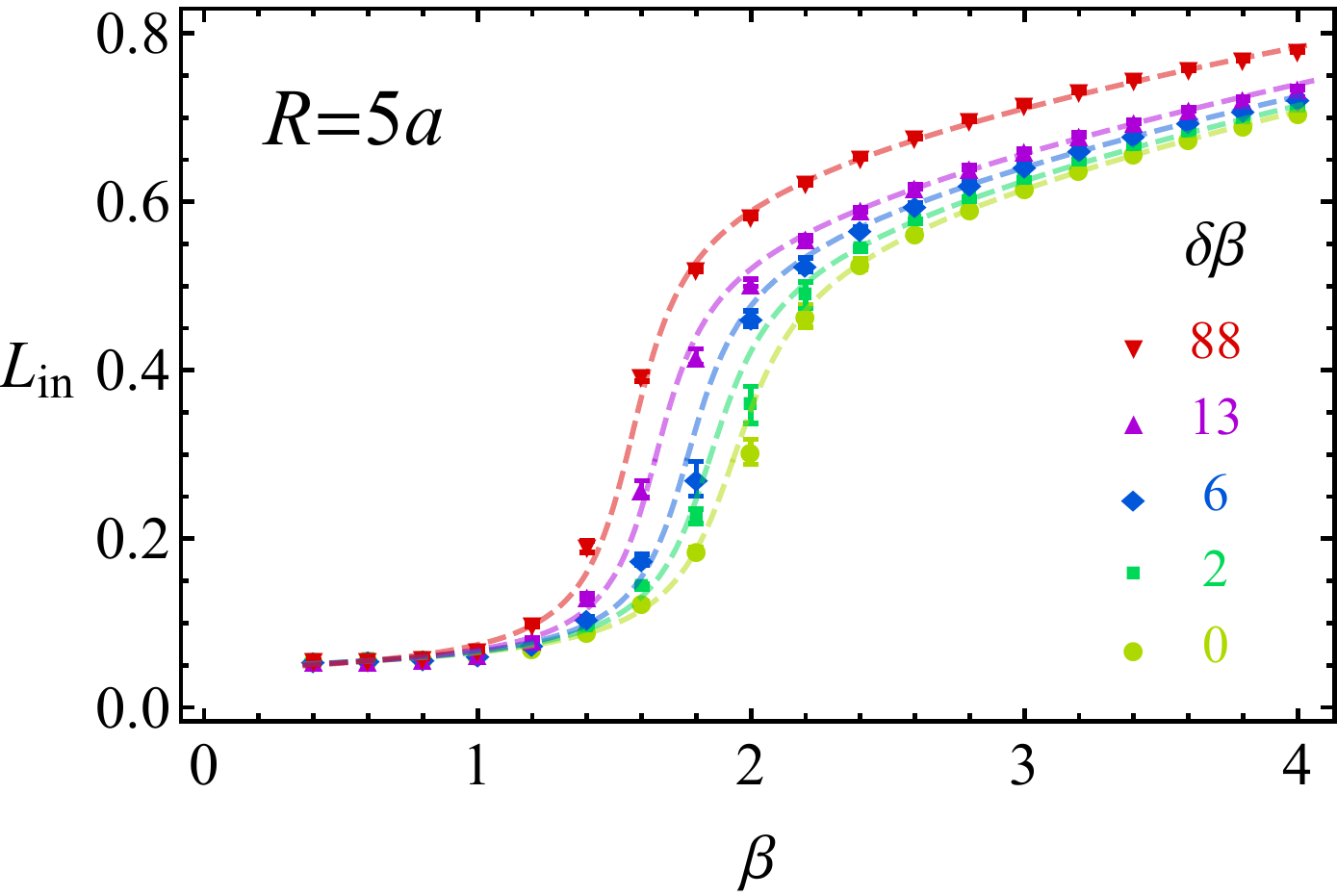} 
\end{center}
\vskip -2mm 
\caption{The expectation value of the Polyakov loop in between the plates~\eq{eq:L:in} vs. the coupling constant $\beta$ at various values of the excess of the coupling constant $\delta \beta$ at the plates~\eq{eq:delta:beta} for  two fixed separations between the plates, $R=2a$ (top) and $R=5a$ (bottom). The curves correspond to the best fits of the data with the fitting function~\eq{eq:fit:L}.}
\label{fig:Polyakov:loop}
\end{figure}

In Fig.~\ref{fig:Polyakov:loop} we show the expectation value of the Polyakov line in the space between the wires for two fixed interwire distances, $R=2a$ and $R=5a$, and for a set of strengths of the Casimir wires $\delta\beta$. We notice a few interesting qualitative properties of the Polyakov line.

Firstly, the Polyakov line $L_{\mathrm{in}}$ in between the wires is a monotonically increasing function of the lattice gauge constant $\beta$. According to Eq.~\eq{eq:T:phys} this fact implies that the Polyakov line increases with the increase in temperature. In turn, the latter property indicates the natural fact that -- similarly to the vacuum in the absence of the Casimir wires -- the deconfinement phase in between the wires is realized at high temperatures while the confinement of electric charges takes place in a low temperature regime. 

Secondly, at a fixed distance between the wires the Polyakov line increases with the increase of the wire strength $\delta \beta$. Therefore, given the relation between the expectation value of the Polyakov line and the free energy of a single charge~\eq{eq:Fx:definition}, we conclude that closely spaced Casimir wires force the vacuum to turn into the deconfinement phase.  This observation agrees with yet another property which can also be seen by comparison of the top and bottom panels of Fig.~\ref{fig:Polyakov:loop}: the closer separation between the wires the higher expectation value of the Polyakov loop. Therefore we come to the conclusions that 
\begin{itemize}
\item[(i)] the Casimir wires tend to make the vacuum between them deconfining;
\item[(ii)] the deconfining influence of the wires becomes stron\-ger as the distance between the wires gets smaller. 
\end{itemize}

Finally, in the low temperature regime the expectation value of the Polyakov loop in the space between the plates $L_{\mathrm{in}}$ does not vanish completely, indicating the absence of a ``pure'' deconfinement. The nonvanishing behavior of an order parameter is a typical property of finite-volume systems. Although in a finite volume the free energy of a single particle does not become infinite, it becomes sufficiently large to mark the appropriate phase as statistically deconfining at $F_{\bs x} \gg T$. In order to characterize the deconfinement of electric charges between the wires, we fit the Polyakov loop by the function
\beqn
L^{\mathrm{fit}}(\beta) = L_0 + L_1 \beta^\nu \Bigl( \arctan [\kappa (\beta - \beta_c)] + \frac{\pi}{2}\Bigr)\,,
\label{eq:fit:L}
\eeqn
where $L_0$, $L_1$, $\nu$, $\kappa$  and $\beta_c$ are fitting parameters. We found that the function~\eq{eq:fit:L} describes our data very well for all available combinations of the interwire distance $R$, the coupling constant $\beta$, and the wire strength $\delta\beta$. Examples of the fits of the numerically obtained Polyakov loop by the analytical function~\eq{eq:fit:L} are shown in Fig.~\ref{fig:Polyakov:loop} by the solid lines.

For all fits the best fit values for the power $\nu$ lie in the range $\nu \approx (1/3 \dots 1/2)$ while the slope parameter $\kappa$ is slowly varying, depending on the interwire distance $R$, in the region $\kappa \approx (4 \dots 8)$. Figure~\ref{fig:Polyakov:loop} suggests that the strength of the deconfinement phase transition gets slightly stronger with the increase of the wire strength $\delta \beta$, while the transition insignificantly weakens with the decrease of the distance between the wires.

The fitting parameter $\beta_c$ plays a role of a pseudocritical value of the lattice coupling constant $\beta = \beta_c$, corresponding to a point where the deconfinement (phase) transition of the finite-volume system takes place. The coupling $\beta_c$ is called the ``pseudocritical'' coupling because in a finite volume the phase transition is, literally speaking, absent and the value of a critical coupling depends on a particular way how it is defined. For example, in Eq.~\eq{eq:fit:L} we could also use a similar function, $f_2(x)=\tanh x + 1$, instead of $f_1(x)=\arctan x + \pi/2$, with a slightly worser quality of the fit. In an infinite-volume limit the pseudocritical coupling constants tend to the same value of the critical coupling. In the next section we will discuss the behavior of the pseudocritical coupling $\beta_c$ defined with the help of Eq.~\eq{eq:fit:L}.

\begin{figure}[!thb]
\begin{center}
\vskip 3mm
\includegraphics[scale=0.55,clip=true]{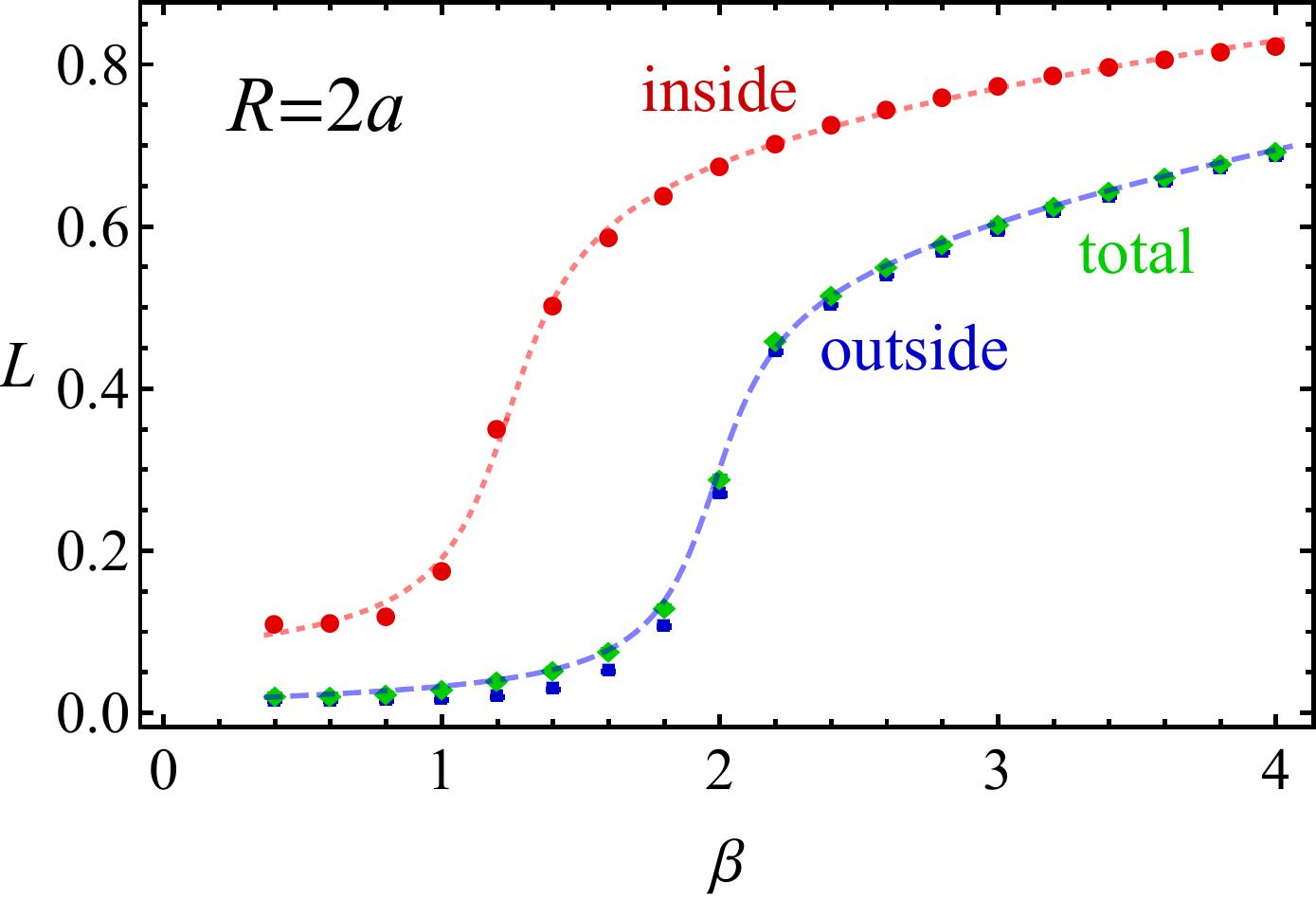}
\end{center}
\vskip -2mm 
\caption{The expectation values of the Polyakov loop: in the whole space~\eq{eq:L:tot} as well as inside~\eq{eq:L:in} and outside~\eq{eq:L:tot} of the wires for a large value of the wire strength $\delta \beta = 88$ with the next-to-minimal separation between the wires, $R = 2 a$.}
\label{fig:loop:compare}
\end{figure}

It is also interesting to compare the relative values of the Polyakov lines in the whole space, as well as inside and outside the wires. In Fig.~\ref{fig:loop:compare} we show the expectation values of the Polyakov line in all three regions, as defined, respectively, by Eqs.~\eq{eq:L:tot}, \eq{eq:L:in}, and \eq{eq:L:out}. In order to maximize the effect we plot the data for the shortest possible distance between the wires ($R=2a$) and at largest available value of the wire strength ($\delta \beta = 88$). As expected, the effect of the wires on phase structure on the vacuum at the exterior of the wires is negligible: the bulk space outside the remains unaffected by the presence of the wires. Using a fit by Eq.~\eq{eq:fit:L} we determine the pseudocritical lattice coupling of the deconfinement transition in the absence of the wires: 
\beqn
\beta_c = 1.97(3)\,.
\label{eq:beta:c:no:wires}
\eeqn
This results agrees within 10\% with an analytical prediction of Ref.~\cite{Chernodub:2001ws} based on properties of lattice monopoles.

In addition, Fig.~\ref{fig:loop:compare} indicates that the Polyakov line in the space between the wires becomes much larger compared to its expectation value outside the wires. This property indicates that the wires make the vacuum in between them deconfining. The deconfinement effect agrees perfectly with the features of the monopole picture described in the previous section which demonstrates that shortly--separated wires expel the magnetic monopoles from the space in between them. Since the monopole gas is tightly linked to the confinement property, the monopole suppression between the plates agrees well with the observed deconfinement effect.

\subsection{Phase diagram in the presence of Casimir wires}

We have found that the vacuum structure of the compact electrodynamics is affected by the presence of dielectric/metallic wires: in the space between two parallel wires the monopole density is diminished while the Polyakov line is, consistently, enhanced. These properties indicate that the parallel wires induce the deconfinement transition which should naturally be reflected in lowering of the deconfinement transition temperature in the volume between the wires. In lattice variables, the physical (pseudo)critical temperature of the deconfinement transition is proportional to the corresponding value of the lattice coupling~\eq{eq:T:phys} so that we should expect that in the presence of the wires the (pseudo)critical value of the coupling $\beta$ is diminished compared to its value in the absence of the wires.

\begin{figure}[!thb]
\begin{center}
\vskip 3mm
\includegraphics[scale=0.35,clip=true]{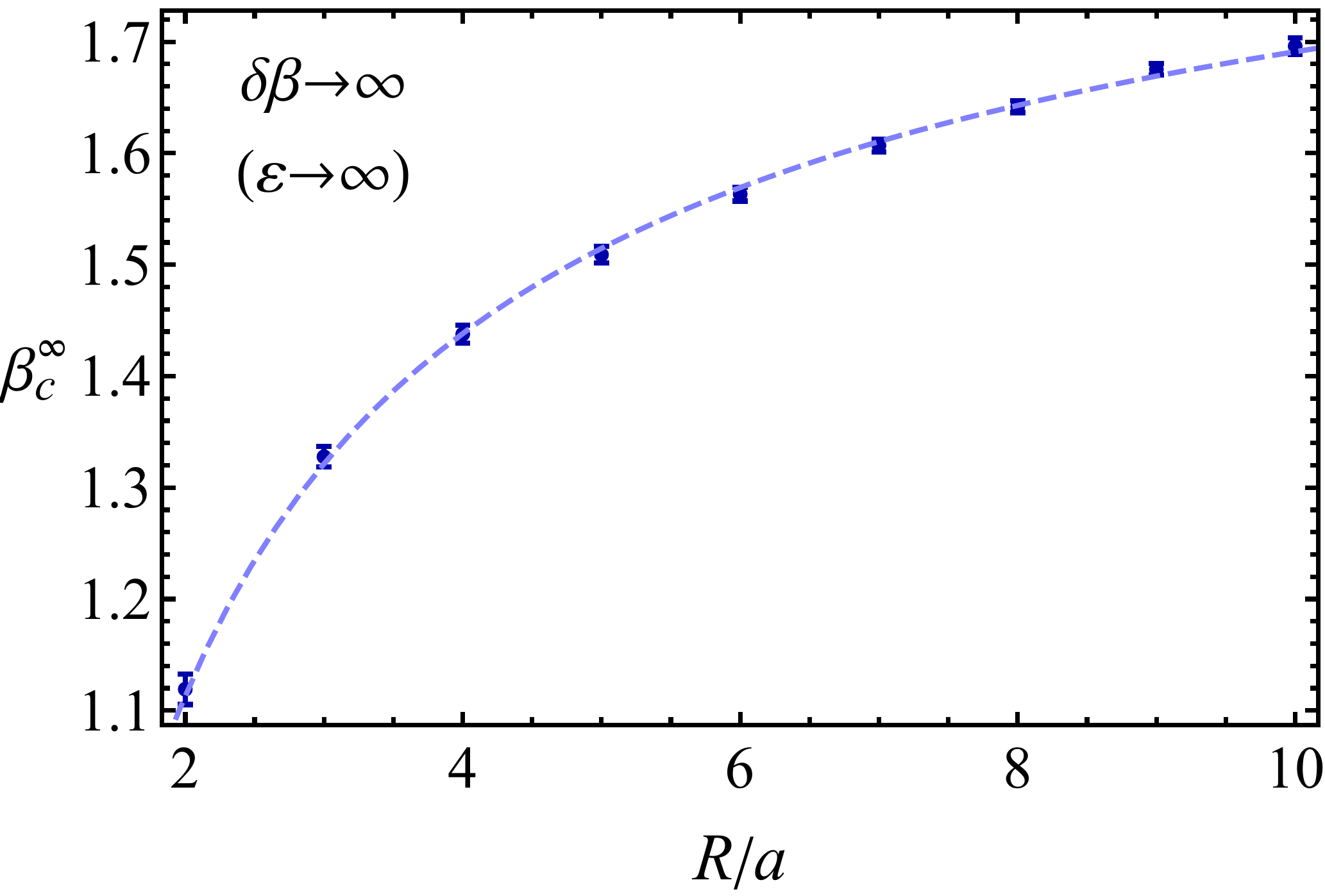}
\end{center}
\vskip -2mm 
\caption{Pseudocritical coupling constant $\beta_c$ of the con\-fi\-ne\-ment-deconfinement transition vs. the interwire distance $R$ for the wires in the metallic limit ($\delta\beta \to \infty$ or $\varepsilon  \to \infty$). The line is drawn to guide eye.}
\label{fig:beta:c:inf}
\end{figure}

In Fig.~\ref{fig:beta:c:inf} we plot the pseudocritical coupling $\beta_c$ determined from the fits of the expectation value of the Polyakov loop by the analytical function~\eq{eq:fit:L}. The coupling $\beta_c$ is shown as a function of the wire strength $\delta\beta$ at a set of various interwire distances~$R$.  Figure~\ref{fig:beta:c:inf} clearly illustrates that the pseudocritical coupling diminishes with the decrease of the distance between the wires. Moreover, the increase of the wire strength $\delta\beta$ -- which is related to the permittivity of the wires~\eq{eq:permittivity:chi} and their dielectric susceptibility~\eq{eq:chi} -- also leads to a substantial lowering of the pseudocritical coupling constant $\beta_c$ for each value of the interwire distance $R$.

\begin{figure}[!thb]
\begin{center}
\vskip 3mm
\includegraphics[scale=0.55,clip=true]{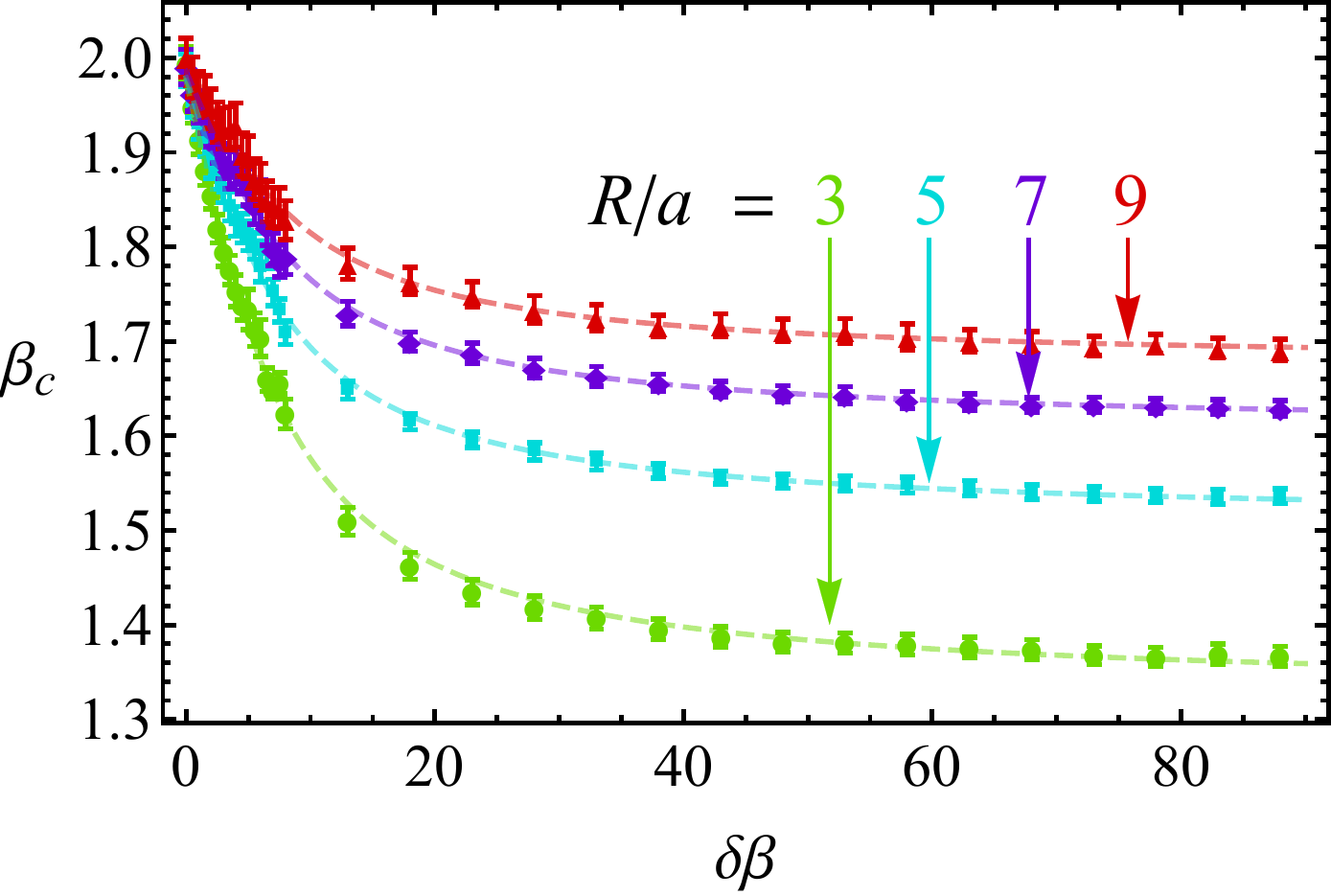}
\end{center}
\vskip -2mm 
\caption{Pseudocritical coupling constant $\beta_c$ of the con\-fi\-ne\-ment-deconfinement transition vs the strength of the Casimir plates $\delta\beta$, Eq.~\eq{eq:delta:beta}, for various values of the separation between the plates $R/a = 3,5,7,9$. The dashed lines are the fits of the corresponding numerical data by Eq.~\eq{eq:fit:beta:c}.}
\label{fig:beta:c}
\end{figure}

It turns out that the dependence of the pseudocritical coupling on the wire strength $\delta\beta$ at each fixed interwire distance $R$ may be very well described by the following simple function:
\beqn
\beta^{\mathrm{fit}}_c(\delta\beta) = \beta^\infty_0 + \beta_1 \left[\tanh (\gamma \, \delta\beta ) - 1\right]\,,
\label{eq:fit:beta:c}
\eeqn
where $\beta^\infty_c = \beta^\infty_c(R)$, $\beta_1 = \beta_1(R)$, and $\gamma = \gamma(R)$ are the fitting parameters which depend on the distance between the wires~$R$. The corresponding fits are shown by the dashed lines in Fig.~\ref{fig:beta:c}.

It is interesting to consider two limits of the fitting function~\eq{eq:fit:beta:c}. In the limit $\delta \beta \to 0$ the wires are absent and the pseudocritical coupling constants for all interwire distances $R$ agree with each other within error bars. Moreover, the statistical average of the pseudocritical couplings for the available set of $R$'s gives us the following averaged coupling of the deconfinement transition:
\beqn
\beta^{(0)}_c \approx 1.988(7).
\eeqn
This value agrees very well with the independent estimate given in Eq.~\eq{eq:beta:c:no:wires}.

In the opposite case we extrapolate, using the fitting function~\eq{eq:fit:beta:c}, the behavior of the pseudocritical coupling constant to the metallic limit where the strength of the wire is infinite, $\delta\beta \to \infty$:
\beqn
\beta^\infty_c  = \lim_{\delta \beta \to \infty} \beta_c(\delta\beta)\,.
\label{eq:beta:c:inf}
\eeqn
In Fig.~\ref{fig:beta:c:inf} we show the extrapolated pseudocritical coupling~\eq{eq:beta:c:inf} as the function of the interwire distance $R$. The coupling $\beta_c^\infty$ is a linearly rising function of the distance between the metallic wires. Using relation~\eq{eq:T:phys} between the physical temperature $T$ and the lattice coupling constant $\beta$, we may easily obtain, from the data of Fig.~\ref{fig:beta:c:inf}, the temperature of the deconfinement transition in the physical units of electromagnetic coupling $g$.

\begin{figure}[!thb]
\begin{center}
\vskip 3mm
\includegraphics[scale=0.55,clip=true]{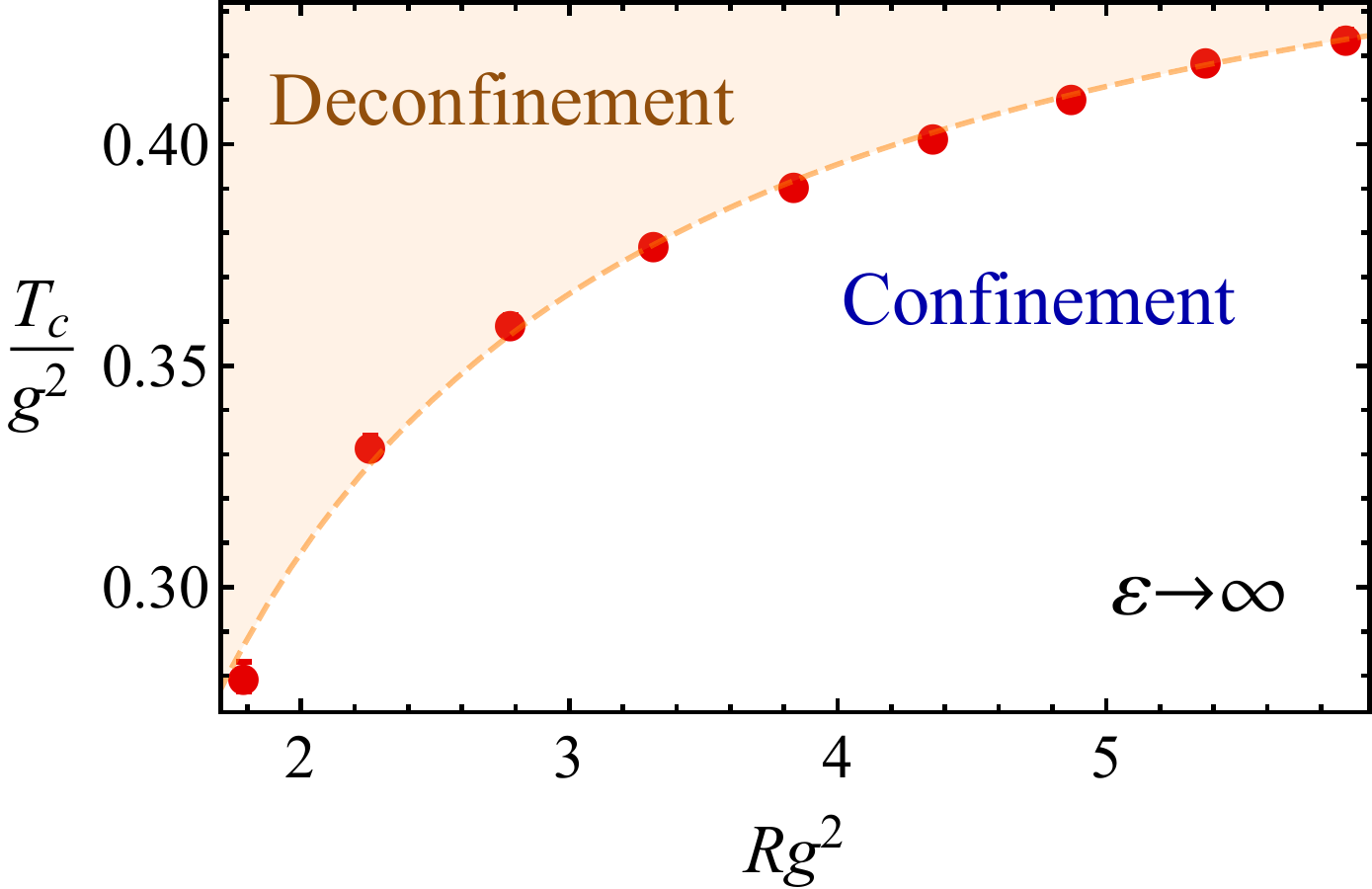}
\end{center}
\vskip -2mm 
\caption{The critical temperature $T_c$ of the confinement-deconfinement transition as the function of the separation between the plates $R$ in the ideal-metal limit ($\varepsilon \to \infty$) in physical units. The dashed line represents the fit by the function~\eq{eq:Tc:R} of the numerical data (the red circles). }
\label{fig:T:c:phys}
\end{figure}

In Fig.~\ref{fig:T:c:phys} we show the phase structure of the vacuum of compact electrodynamics in the space between the infinitely long parallel Casimir wires in the metallic limit $\varepsilon \to \infty$. The (pseudo)critical temperature $T_c$ is a monotonically rising function of the inter-wire distance~$R$. It is remarkable that the dependence of the deconfinement temperature on the interwire distance $R$ in the available fitting range $R g^2 \in (1.8,6.0)$ may be well described by the following simple function:
\beqn
T_c(R) = T_c^\infty - \frac{C_0}{R}\,,
\label{eq:Tc:R}
\eeqn
where the best fit parameter $C_0 = 0.35(1)$ determines the slope  of the dependence of the pseudocritical temperature $T_c(R)$ on the interwire distance~$R$. In the limit of the infinitely separated wires the critical temperature is given by the following extrapolation of~\eq{eq:Tc:R}:
\beqn
T_c^\infty \equiv \lim_{R \to \infty} T_c(R) = 0.483(2) \cdot g^2\,.
\label{eq:Tc:fits}
\eeqn
The best fit function~\eq{eq:Tc:R} with the mentioned parameters $C_0$ and $T_c^\infty$ is shown in Fig.~\ref{fig:T:c:phys} by the dashed line.

Thus, the closely spaced wires affect the vacuum structure of the compact electrodynamics by inducing, in the space between them, the deconfinement of electric charges.  Formally, the confinement property disappears completely when the separation between the plates becomes smaller ($R \leqslant R_c$)  than the critical radius $R=R_c$ at which the critical temperature vanishes:
\beqn
T_c(R_c) = 0\,.
\label{eq:Tc:R:0}
\eeqn
According to Eqs.~\eq{eq:Tc:R} and \eq{eq:Tc:fits} the critical distance between the metallic wires ($\varepsilon \to \infty$) can be estimated with the help of Eq.~\eq{eq:Tc:R:0} as follows:
\beqn
R_c = 0.72(1) \, \frac{1}{g^2}\,.
\label{eq:R:c}
\eeqn
We stress that the estimate~\eq{eq:R:c} is based on the assumption that the fitting function~\eq{eq:Tc:R} remains valid in the region of small interwire distances which were not discussed in our paper.

\section{Conclusions}

In our paper we have shown that the Casimir effect leads to deconfinement of electric charges via modification of quantum fluctuations in a confining field theory. 

In order to illustrate this statement we have studied the compact lattice electrodynamics (compact QED) from the first principles of the theory using numerical methods developed in our earlier papers~\cite{ref:paper:1,ref:paper:2}. The compact QED in two space dimensions serves as a toy model of the more complicated theory of strong interactions, QCD, as both these theories possess the phenomena of the linear confinement of charges and the mass gap generation. At finite temperature the vacua of both theories exhibit a phase transition from the confining phase to the deconfining phase. Therefore, the compact QED in (2+1) spacetime dimensions is a useful toy model to study the effects of the Casimir geometry on confining properties of the vacuum of the theory. Due to the reduced spatial dimensionality  in (2+1) dimensions it is appropriate to formulate the Casimir problem between the wires rather than between the plates. 

In our first study~\cite{ref:paper:1} we have demonstrated the reliability of the new numerical approach based on first-principle simulations of lattice (gauge) theories. We have confirmed that our numerical method correctly reproduces the known analytical results for the Casimir interaction between the dielectric and metallic wires in a weakly coupled region of compact electrodynamics where the (nonperturbative) interaction effects are small. 

In our second study~\cite{ref:paper:2} we used the same approach to reveal nonperturbative effects of magnetic monopoles on the Casimir energy in the strong coupling region. We have found that the virtual monopoles and antimonopoles make the Casimir interaction short-ranged while increasing the attractive Casimir-Polder force at short distances. 

In the present, third paper we demonstrate that while both perturbative~\cite{ref:paper:1} and nonperturbative~\cite{ref:paper:2} vacuum effects modify the Casimir energy between the physical wires, the Casimir effect, in the other way around, causes the restructuring of the nonperturbative vacuum in the space between the wires. We studied the theory at finite temperature in order to highlight the deconfining effect of the shortly-spaced wires on the vacuum. We also investigated the wires of different ``wire strength'' $\delta\beta$ which is related to the permittivity of the wires~\eq{eq:permittivity:chi} and their dielectric susceptibility~\eq{eq:chi}.

\begin{figure}[!thb]
\begin{center}
\vskip 5mm
\includegraphics[scale=0.11,clip=true]{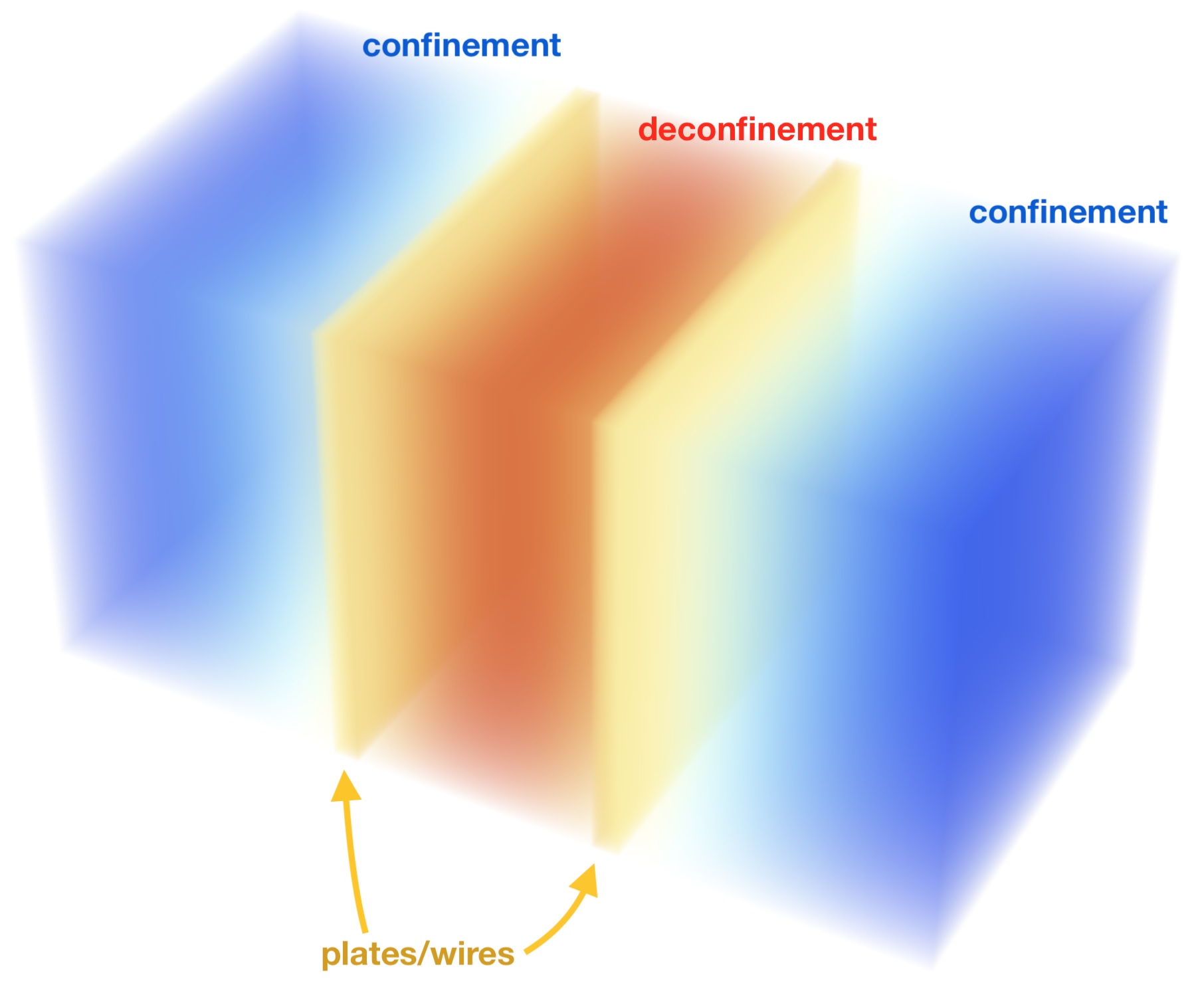}
\end{center}
\vskip -2mm 
\caption{Phase transition due to the Casimir effect: closely spaced wires (plates) lead to deconfinement of electric charges in the confining phase of compact lattice electrodynamics.}
\label{fig:illustration}
\end{figure}

The effect is illustrated in Fig.~\ref{fig:illustration}: the vacuum in the space between the Casimir wires becomes deconfining as the distance between the wires gets smaller. In (2+1) dimensions this figure represents two spatial and one Euclidean time dimension. The dependence of the deconfining temperature $T_c$ on the distance $R$ between perfectly metallic wires is shown, in physical units, in Fig.~\ref{fig:T:c:phys}. The critical temperature is very well described by the simple function give in Eq.~\eq{eq:Tc:R} suggesting that at certain critical distance between the wires~\eq{eq:R:c} the vacuum in the space between the wires resides always in the deconfinement phase regardless of the temperature of the system. In a low-temperature regime the vacuum outside the plates always stays in the confinement phase. We suggest that the same Casimir-induced confinement-suppression effect is also realized in 3+1 dimensional confining theories (in this case Fig.~\ref{fig:illustration} illustrates the Casimir plates in three spatial dimensions). 

We conclude that the dynamical monopoles, linear confinement and mass gas generation on the one side and the Casimir geometry on the other side are antagonistic to each other: the presence of the dynamical monopoles in the vacuum suppresses the Casimir-Polder force at large distances via the mass gap generation~\cite{ref:paper:2}, while, in the other way around, the dielectric and/or metallic wires in the Casimir problem diminish the mean monopole density thus leading to the absence of the linear confinement of electric charges at large distances and to reduction of the mass gap for photons in the vacuum between the wires.

\acknowledgments

The research was carried out within the state assignment of the Ministry of Science and Education of Russia (Grant No. 3.6261.2017/8.9). The numerical simulations were performed at the computing cluster Vostok-1 of Far Eastern Federal University.


\begin{thebibliography}{99}

\bibitem{ref:Casimir}
H.B.G Casimir, 
``On the attraction between two perfectly conducting plates'',
Proc. Kon. Ned. Akad. Wetensch. {\bf 51} (1948), 793.

\bibitem{ref:Bogdag}
M. Bordag, G. L. Klimchitskaya, U. Mohideen, and V. M. Mostepanenko, 
``Advances in the Casimir Effect'' (Oxford University Press, New York, 2009).

\bibitem{ref:Milton}
K. A. Milton,
``The Casimir Effect: Physical Manifestations of Zero-Point Energy''
(World Scientific Publishing, Singapore, 2001).

\bibitem{Casmir:1947hx} 
  H.~B.~G.~Casimir and D.~Polder,
  ``The Influence of retardation on the London-van der Waals forces,''
  Phys.\ Rev.\  {\bf 73}, 360 (1948).
%  doi:10.1103/PhysRev.73.360
  %%CITATION = doi:10.1103/PhysRev.73.360;%%

\bibitem{ref:experiment}
S. K. Lamoreaux, ``Demonstration of the Casimir Force in the 0.6 to 6$\,\mu$m Range'',
Phys. Rev. Lett. 78, 5 (1997) [Erratum ibid, 5475 (1998)];
U. Mohideen and A.~Roy
``Precision Measurement of the Casimir Force from 0.1 to  0.9$\,\mu$m'',
Phys. Rev. Lett. 81, 4549 (1998);
G. Bressi, G. Carugno, R. Onofrio, and G. Ruoso,
``Measurement of the Casimir Force between Parallel Metallic Surfaces'',
Phys. Rev. Lett. 88, 041804 (2002).

\bibitem{ref:proximity}
B. V. Derjaguin,  I. I. Abrikosova  and  E. M. Lifshitz, 
``Direct measurement of molecular attraction between solids separated by a narrow gap'', Q. Rev. Chem. Soc. 10, 295 (1956);
J. Blocki, J. Randrup, W.J. Swiatecki, C.F. Tsang, ``Proximity forces'', Ann. Phys. (N.Y.) 105, 427 (1977).

\bibitem{Johnson:2010ug} 
  S.~G.~Johnson,
  ``Numerical methods for computing Casimir interactions,''
  Lect.\ Notes Phys.\  {\bf 834}, 175 (2011).
%  doi:10.1007/978-3-642-20288-9_6
%  [arXiv:1007.0966 [quant-ph]].
  %%CITATION = doi:10.1007/978-3-642-20288-9_6;%%

\bibitem{Gies:2006cq} 
  H.~Gies and K.~Klingmuller,
  ``Worldline algorithms for Casimir configurations,''
  Phys.\ Rev.\ D {\bf 74}, 045002 (2006)
 % doi:10.1103/PhysRevD.74.045002
 [quant-ph/0605141];
  %%CITATION = doi:10.1103/PhysRevD.74.045002;%%
  H.~Gies, K.~Langfeld and L.~Moyaerts,
  ``Casimir effect on the worldline,''
  JHEP {\bf 0306}, 018 (2003)
%  doi:10.1088/1126-6708/2003/06/018
 [hep-th/0303264].
  %%CITATION = doi:10.1088/1126-6708/2003/06/018;%%  

\bibitem{ref:Oleg} 
  O.~Pavlovsky and M.~Ulybyshev,
  ``Casimir energy in the compact QED on the lattice,''
  arXiv:0901.1960;
%%CITATION = ARXIV:0901.1960;%%
%  O.~Pavlovsky and M.~Ulybyshev,
  ``Casimir energy calculations within the formalism of the noncompact lattice QED,''
  Int.\ J.\ Mod.\ Phys.\ A {\bf 25}, 2457 (2010) [arXiv:0911.2635 [hep-lat]];
  %%CITATION = doi:10.1142/S0217751X10048378;%%
  ``Casimir energy in noncompact lattice electrodynamics,''
  Theor.\ Math.\ Phys.\  {\bf 164}, 1051 (2010);
  %%CITATION = doi:10.1007/s11232-010-0084-5;%%
%  O.~Pavlovsky and M.~Ulybyshev,
  ``Monte-Carlo calculation of the lateral Casimir forces between rectangular gratings within the formalism of lattice quantum field theory,''
  Int.\ J.\ Mod.\ Phys.\ A {\bf 26}, 2743 (2011) [arXiv:1105.0544 [quant-ph]].
  %%CITATION = doi:10.1142/S0217751X11053559;%%
  
\bibitem{ref:paper:1} 
 M.~N.~Chernodub, V.~A.~Goy and A.~V.~Molochkov,
  ``Casimir effect on the lattice: U(1) gauge theory in two spatial dimensions,''
  Phys.\ Rev.\ D {\bf 94}, no. 9, 094504 (2016)
  %doi:10.1103/PhysRevD.94.094504
  [arXiv:1609.02323 [hep-lat]].
  %%CITATION = doi:10.1103/PhysRevD.94.094504;%%
  
\bibitem{ref:paper:2} 
  M.~N.~Chernodub, V.~A.~Goy and A.~V.~Molochkov,
  ``Nonperturbative Casimir effect and monopoles: compact Abelian gauge theory in two spatial dimensions,''
  Phys.\ Rev.\ D {\bf 95}, no. 7, 074511 (2017)
  %doi:10.1103/PhysRevD.95.074511
  [arXiv:1703.03439 [hep-lat]].
  %%CITATION = doi:10.1103/PhysRevD.95.074511;%%

\bibitem{Bordag:1983zk} 
  M.~Bordag, D.~Robaschik and E.~Wieczorek,
  ``Quantum field theoretic treatment of the Casimir effect,''
  Annals Phys.\  {\bf 165}, 192 (1985).
%  doi:10.1016/S0003-4916(85)80009-9
  %%CITATION = doi:10.1016/S0003-4916(85)80009-9;%%

\bibitem{Flachi:2013bc} 
  A.~Flachi,
``Strongly Interacting Fermions and Phases of the Casimir Effect,''
  Phys.\ Rev.\ Lett.\  {\bf 110}, no. 6, 060401 (2013)
%  doi:10.1103/PhysRevLett.110.060401
  [arXiv:1301.1193 [hep-th]].
  %%CITATION = doi:10.1103/PhysRevLett.110.060401;%%

\bibitem{Tiburzi:2013vza} 
  B.~C.~Tiburzi,
  ``Chiral Symmetry Restoration from a Boundary,''
  Phys.\ Rev.\ D {\bf 88}, 034027 (2013)
%  doi:10.1103/PhysRevD.88.034027
  [arXiv:1302.6645 [hep-lat]].
  %%CITATION = doi:10.1103/PhysRevD.88.034027;%%

\bibitem{Chernodub:2016kxh} 
  M.~N.~Chernodub and S.~Gongyo,
  ``Interacting fermions in rotation: chiral symmetry restoration, moment of inertia and thermodynamics,''
  JHEP {\bf 1701}, 136 (2017)
%  doi:10.1007/JHEP01(2017)136
  [arXiv:1611.02598 [hep-th]];
  %%CITATION = doi:10.1007/JHEP01(2017)136;%%
%\cite{Chernodub:2017ref}
%\bibitem{Chernodub:2017ref} 
%  M.~N.~Chernodub and S.~Gongyo,
  ``Effects of rotation and boundaries on chiral symmetry breaking of relativistic fermions,''
  Phys.\ Rev.\ D {\bf 95}, no. 9, 096006 (2017)
%  doi:10.1103/PhysRevD.95.096006
  [arXiv:1702.08266 [hep-th]].
  %%CITATION = doi:10.1103/PhysRevD.95.096006;%%

\bibitem{Flachi:2017cdo} 
  A.~Flachi, M.~Nitta, S.~Takada and R.~Yoshii,
 ``Sign Flip in the Casimir Force for Interacting Fermion Systems,''
  Phys.\ Rev.\ Lett.\  {\bf 119}, no. 3, 031601 (2017)
%  doi:10.1103/PhysRevLett.119.031601
  [arXiv:1704.04918 [hep-th]].
  %%CITATION = doi:10.1103/PhysRevLett.119.031601;%%

\bibitem{Flachi:2017xat} 
  A.~Flachi, M.~Nitta, S.~Takada and R.~Yoshii,
  ``Casimir Force for the ${\mathbb C}P^{N-1}$ Model,''
  arXiv:1708.08807 [hep-th].
  %%CITATION = ARXIV:1708.08807;%%

\bibitem{Betti:2017zcm} 
  A.~Betti, S.~Bolognesi, S.~B.~Gudnason, K.~Konishi and K.~Ohashi,
  ``Large-N CP(N-1) sigma model on a finite interval and the renormalized string energy,''
  arXiv:1708.08805 [hep-th].
  %%CITATION = ARXIV:1708.08805;%%
 
\bibitem{tHooft:1974kcl} 
  G.~'t Hooft,
``Magnetic Monopoles in Unified Gauge Theories,''
  Nucl.\ Phys.\ B {\bf 79}, 276 (1974).
%  doi:10.1016/0550-3213(74)90486-6
  %%CITATION = doi:10.1016/0550-3213(74)90486-6;%%

\bibitem{Polyakov:1974ek} 
  A.~M.~Polyakov,
``Particle Spectrum in the Quantum Field Theory,''
  JETP Lett.\  {\bf 20}, 194 (1974).
%  [Pisma Zh.\ Eksp.\ Teor.\ Fiz.\  {\bf 20}, 430 (1974)].
  %%CITATION = JTPLA,20,194;%%

\bibitem{Georgi:1974sy} 
  H.~Georgi and S.~L.~Glashow,
  ``Unity of All Elementary Particle Forces,''
  Phys.\ Rev.\ Lett.\  {\bf 32}, 438 (1974).
%  doi:10.1103/PhysRevLett.32.438
  %%CITATION = doi:10.1103/PhysRevLett.32.438;%%
  
\bibitem{ref:book:Herbut}
I. Herbut, 
``A Modern Approach to Critical Phenomena'' (Cambridge University Press, 2007). 

\bibitem{ref:book:Kleinert}
Hagen Kleinert, 
Multivalued Fields: In Condensed Matter, Electromagnetism, and Gravitation (World Scientific, 2008, Singapore).

\bibitem{Polyakov:1976fu} 
  A.~M.~Polyakov,
  ``Quark Confinement and Topology of Gauge Groups,''
  Nucl.\ Phys.\ B {\bf 120}, 429 (1977).
%  doi:10.1016/0550-3213(77)90086-4
  %%CITATION = doi:10.1016/0550-3213(77)90086-4;%%

\bibitem{Chernodub:1997ay} 
  M.~N.~Chernodub and M.~I.~Polikarpov,
  ``Abelian projections and monopoles,''
  In *Cambridge 1997, Confinement, duality, and nonperturbative aspects of QCD* 387 
  [hep-th/9710205].
%%CITATION = HEP-TH/9710205;%%

\bibitem{ref:Gattringer} 
C. Gattringer, C.B. Lang, ``Quantum Chromodynamics on the Lattice'' (Springer-Verlag, Berlin Heidelberg, 2010).

\bibitem{ref:Omelyan}
I. P. Omelyan, I. M. Mryglod, and R. Folk, 
``Optimized Verlet-like algorithms for molecular dynamics simulations'', Phys. Rev. E {\bf 65}, 056706 (2002) [cond-mat/0110438 [cond-mat.stat-mech]];
%I. P. Omelyan, I. M. Mryglod, and R. Folk,
``Symplectic analytically integrable decomposition algorithms: classification, derivation, and application to molecular dynamics, quantum and celestial mechanics simulations'', Comput. Phys. Commun. {\bf 151}, 272 (2003).

\bibitem{ref:Sexton}
  J.~C.~Sexton and D.~H.~Weingarten,
  ``Hamiltonian evolution for the hybrid Monte Carlo algorithm,''
  Nucl.\ Phys.\ B {\bf 380}, 665 (1992);
%  doi:10.1016/0550-3213(92)90263-B
  %%CITATION = doi:10.1016/0550-3213(92)90263-B;%%
%\cite{Urbach:2005ji}
%\bibitem{Urbach:2005ji} 
  C.~Urbach, K.~Jansen, A.~Shindler and U.~Wenger,
  ``Hybrid Monte Carlo algorithm with multiple time scale integration and mass preconditioning,''
  Comput.\ Phys.\ Commun.\  {\bf 174}, 87 (2006)
%  doi:10.1016/j.cpc.2005.08.006
  [hep-lat/0506011].
  %%CITATION = doi:10.1016/j.cpc.2005.08.006;%%
  
\bibitem{Chernodub:2001ws} 
  M.~N.~Chernodub, E.~M.~Ilgenfritz and A.~Schiller,
  ``A Lattice study of 3-D compact QED at finite temperature,''
  Phys.\ Rev.\ D {\bf 64}, 054507 (2001)
%  doi:10.1103/PhysRevD.64.054507
 [hep-lat/0105021].
  %%CITATION = doi:10.1103/PhysRevD.64.054507;%%
  
\end{thebibliography}
\end{document}